\documentclass[%
reprint,
superscriptaddress,
amsmath,amssymb,
aps,
prmaterials,
]{revtex4-2}

\usepackage{graphicx}% Include figure files
\usepackage{dcolumn}% Align table columns on decimal point
\usepackage{bm}% bold math
\usepackage{hyperref}% add hypertext capabilities
\usepackage{cleveref}
\usepackage{siunitx}
\usepackage{mathtools}
\usepackage[caption=false]{subfig} 
\usepackage{enumitem}
\usepackage{multirow}
\usepackage{scalerel,amssymb}
\usepackage{verbatim}
\usepackage[dvipsnames]{xcolor}
\usepackage{lineno}

\usepackage[version=4]{mhchem}
\UseRawInputEncoding

\crefname{figure}{Figure}{Figures} % Standard reference
\Crefname{figure}{Figure}{Figures} % Beginning-of-sentence ref
\crefname{table}{Table}{Tables}
\Crefname{table}{Table}{Tables}
\crefname{equation}{Eq.}{Eq.}
\Crefname{equation}{Equation}{Equations}
\begin{document}
\title{Design guidelines for two-dimensional transition metal dichalcogenide alloys}

\author{Andrea Silva}
\email{a.silva@soton.ac.uk}
\affiliation{Faculty of Engineering and Physical Sciences, University of Southampton, University Road, SO17 1BJ Southampton, United Kingdom}
\affiliation{national Centre for Advanced Tribology Study, University Road, SO17 1BJ Southampton, United Kingdom}

\author{Jiangming Cao}
\affiliation{Faculty of Mechanical and Civil Engineering, Helmut-Schmidt-Univeristy, Holstenhofweg 85, 22043 Hamburg, Germany}

\author{Tomas Polcar}
\affiliation{Faculty of Engineering and Physical Sciences, University of Southampton, University Road, SO17 1BJ Southampton, United Kingdom}
\affiliation{Advanced Materials Group, Faculty of Electrical Engineering, Czech Technical University in Prague (CTU), Karlovo Náměstí 13, 12135 Prague, Czech Republic}

\author{Denis Kramer}
\email{d.kramer@hsu-hh.de}
\affiliation{Faculty of Engineering and Physical Sciences, University of Southampton, University Road, SO17 1BJ Southampton, United Kingdom}
\affiliation{Faculty of Mechanical and Civil Engineering, Helmut-Schmidt-Univeristy, Holstenhofweg 85, 22043 Hamburg, Germany}
\affiliation{Department of Heterogeneous Catalysis, Helmholtz-Zentrum Hereon, Max-Planck-Strasse 1, 21502 Geesthacht, Germany}


%\begin{abstract}    
%\textbf{Abstract}
%\\
%Two-dimensional (2D) materials and Transition Metal Dichalcogenides (TMD) in particular are at the forefront of nanotechnology. 
%To tailor properties for engineering applications, alloying strategies used for bulk metals in the last century need to be extended to this novel class of materials.
%Here we present a systematic analysis of the phase behaviour of substitutional 2D alloys in the TMD family on both the metal and chalcogenide site.
%
%The phase behaviour is quantified in terms of a metastability metric and benchmarked against systematic computational screening of configurational energy landscapes.
%
%The resulting Pettifor maps can be used to identify broad trends across chemical spaces and as starting point for setting up rational search strategies in phase space, thus allowing for targeted computational analysis of properties on likely thermodynamically  stable compounds.
%
%The results presented here also constitute a useful guideline for synthesis of binary metal 2D TMDs alloys via a range of synthesis techniques.
%\end{abstract}

\maketitle

\section*{Abstract}
Two-dimensional (2D) materials and Transition Metal Dichalcogenides (TMD) in particular are at the forefront of nanotechnology. 
To tailor their properties for engineering applications, alloying strategies---used successfully for bulk metals in the last century---need to be extended to this novel class of materials.
Here we present a systematic analysis of the phase behaviour of substitutional 2D alloys in the TMD family on both the metal and chalcogenide site.
The phase behaviour is quantified in terms of a metastability metric and benchmarked against systematic computational screening of configurational energy landscapes from {\it First Principles}.
The resulting Pettifor maps can be used to identify broad trends across chemical spaces and as starting point for setting up rational search strategies in phase space, thus allowing for targeted computational analysis of properties on likely thermodynamically  stable compounds.
The results presented here also constitute a useful guideline for synthesis of binary metal 2D TMDs alloys via a range of synthesis techniques.

%\linenumbers % Put linenumbers
\section{Intro}
Since the discovery of graphene, 2D materials have been a frontier in Materials Science and Discovery.
Their unique properties and reduced dimensionality have sparked an interest in nanoscale engineering applications, in addition to fundamental research interests~\cite{Smolenski2020}.
Ideas for 2D-materials-based devices can be found in tribology~\cite{Song2018RobustSuperlub}, electronics~\cite{Das2015} and  catalysis~\cite{Pattengale2020} amongst other areas.
Up to now, most research efforts have focused on identifying 2D unaries and binaries both theoretically~\cite{Mounet2018,Sorkun2020,Kumar2022TMD_ML} and experimentally~\cite{Zhou2018,Shivayogimath2018} with only  limited attempts to exploit the vast chemical space spanned by alloys to optimise properties.
Therefore, little is known about their thermodynamic phase behaviour.
The structures and orderings of possible alloys are largely unexplored territory~\cite{Domask2015,Woods-Robinson2022alloy_screening}.
A few 2D ternaries have been reported experimentally~\cite{Koepernik2016,SAEKI1987}, but no systematical analysis across chemical spaces has been carried out, although a handful of binary alloy systems have been studied~\cite{Gao2020,Han2020a,Chen2013}.
But knowledge of their thermodynamic properties is fundamental for rationally advancing the engineering applications of 2D materials.
For instance, the presence of miscibility gaps and competing ternaries has to be taken into account when properties such as bandgap and electronic transport are tuned to desired values by chemical doping~\cite{Worsdale2015a}.

Due to superior scalability, computational tools can complement experimental efforts by efficiently scanning phase space to provide guidelines for synthesis and estimates of properties, potentially reducing the number of viable candidates by orders of magnitude.
For example, Mounet \textit{et al.}~\cite{Mounet2018} reduced a dataset of $\SI{1e5}{}$ bulk crystal structures from experimental databases to 258 easily exfoliable monolayer (ML) candidates, which has to be compared with dozens of candidates that are usually the subject of large-scale experimental studies~\cite{Zhou2018,Shivayogimath2018}.

Empirical rules like the Hume-Rothery rules~\cite{Abbott} and bulk Pettifor maps~\cite{Pettifor1986} have guided the discovery of metallic bulk alloys in the last century. The wide validity of these simple rules in metallic alloys is somewhat surprising but has been comprehensively verified by a symbiotic relationship between experiments and simulations.
The phase diagrams of these bulk alloys have been mapped out experimentally since the 1940s, and later integrated with and rationalised with predictive theories enabled by the advent of Density Functional Theory (DFT) and Cluster Expansion (CE) methods~\cite{Connolly1983} in the 1980s.
More recently, empirical rules have been cast in terms of probabilistic models trained on computational datasets~\cite{Hautier2011} or extended to include the physics of oxides~\cite{Ceder2000}. 

Here, we compile a dataset of two-dimensional TMD compounds in different prototypes and explore technologically relevant alloying possibilities on both the metal and chalcogenide sites.
The results allow us to extend the Hume-Rothery rules to this class of materials and build Pettifor maps for substitutional alloys as a visual tool to navigate the chemical space of two-dimensional TMDs.
Selected predictions by this map are benchmarked against DFT calculations and experimental results from the literature, achieving remarkable agreement.

%------------------------------------------------------
% Chem and proto space
%------------------------------------------------------
\section{Chemical and Coordination Spaces}\label{sec:chem_space}
\begin{figure*}[!htb]
    \centering
    \includegraphics[width=\textwidth]{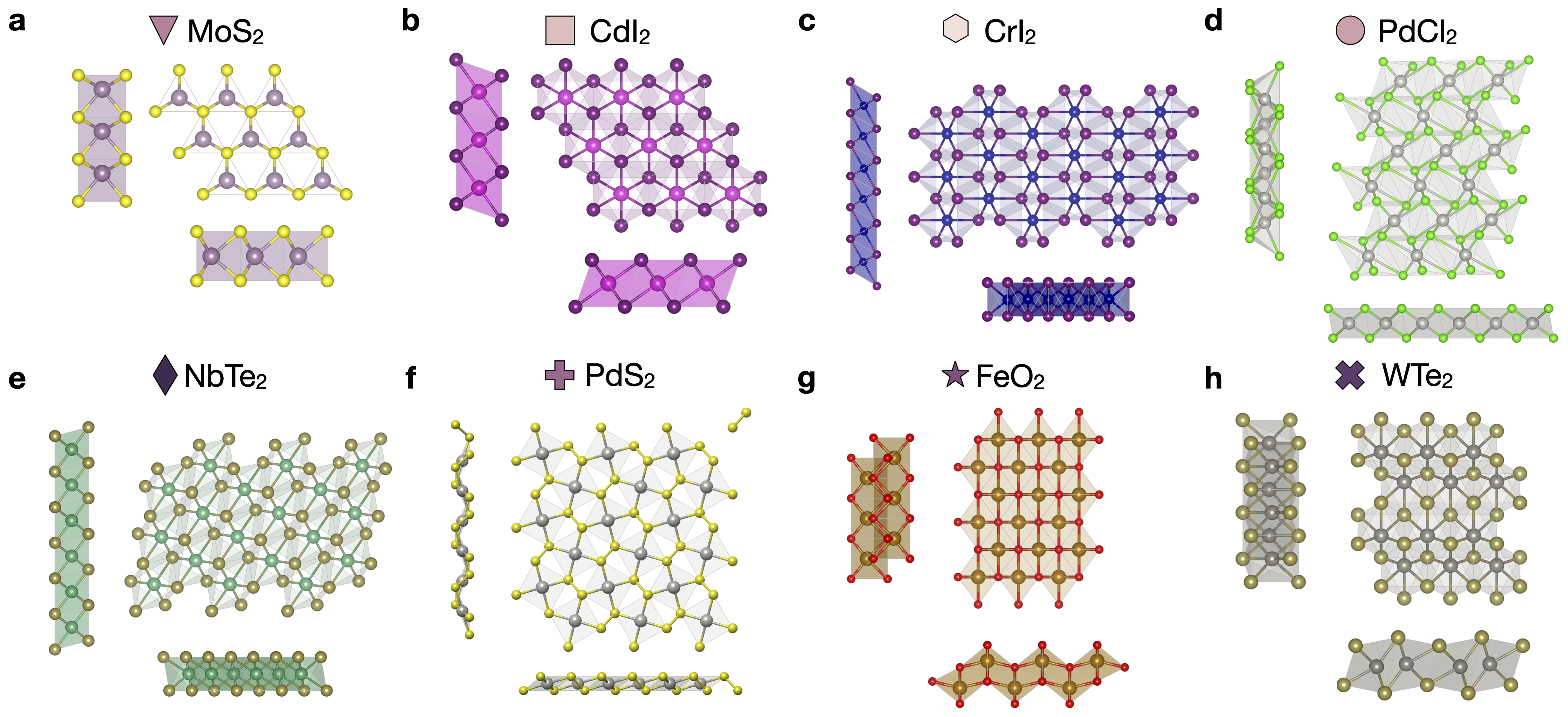}
    \caption[]{
    (a-h) Sides and top views of the eight $MX_2$ prototypes. 
    The prototype will be identify in the rest of the paper by the symbol left of the name.
    The space group of each prototype is reported in Table SIII of the SI.
    }
    \label{fig:p_table-TM_calc}
\end{figure*}

The space of considered structural prototypes for 2D TMD alloys is built from the database compiled by Mounet and coworkers~\cite{Mounet2018}, comprising 258 mechanically stable ML structures identified from experimental bulk compounds.
Thus, the phase stability study is conducted on ML geometries only.
The selection of the possible prototypes and elements to mix is further guided by literature knowledge~\cite{Mounet2018,Furlan2015,Shivayogimath2018,Onofrio2017} to filter the original database according to the class of materials of interest.
The database is scanned for compounds of the form
\ce{MX2}, where \ce{M} is a TM cation and \ce{X} is the anion, oxidising the TM (see Section I in the SI for the list of cations and anions considered).
While selecting the prototypes, the possible cations are restricted to the transition metals considered, but the anions are not limited to chalcogenides, because layered prototypes that could host TMD alloys may not be expressed in terms of chalcogenides in the database (see Section I and table SII in the SI for details).
This search yields the $N_p=8$ prototypes shown in \cref{fig:p_table-TM_calc}a-h, whose space groups are reported in Table SIII of the SI.

Intermediate TMs (Cr, Mn, Fe, Ru, Os) are considered here, although they do not form layered chalcogenides on their own but might form ML alloys in combination with other TMs, e.g. Fe-doped MoS$_2$ ML~\cite{Furlan2015}.
Late transition metals from group XI onward are excluded, as they do not bind with chalcogenides to form layered materials~\cite{Shivayogimath2018}.
This yields $N_{\rm M}=21$ TMs as possible cations \ce{M} in the \ce{MX2} stochiometry, with ${\rm X}=\mathrm{(S, Se, Te)}$ as possible $N_{\rm X}=3$ anions.

While the methodology described here is valid for any stochiometry and cation-anion selection, our analysis will focus on \ce{MX_2} compounds, as these are the most frequently synthesised and studied compounds of the family.
This selection yields $N_{\rm M} \times N_{\rm X} \times N_p = 504$ binaries as a starting point for $N_{\rm M}^2\times N_{\rm X} \times N_p = 10584$ substitutional alloys on the TM site and $N_{\rm X}^2\times N_{\rm M} \times N_p = 1512$ substitutional alloys on the chalcogenide site.
The total number of candidates, although large from an experimental point of view, allows for an exhaustive theoretical analysis rather than approximate methods based on statistical sampling of configurational space~\cite{Avery2019}.
%

%\subsection{Lattice stability}\label{sec:formen_proto}
%
The energy above the ground state of each compound \ce{MX_2} in a given prototype $p$, also known as \textit{lattice stability}, is given by the total energy per site with respect to the ground state (GS)~\cite{Wang2004Calphad,Silva2022pettifor}.
For varying TM and fixed chalcogenide, the lattice stability reads
\begin{align}
\label{eq:MS2_formen}
    E_\mathrm{F}({\rm M}, p) = E({\rm M}, p) - E_\mathrm{GS}({\rm M}),
\end{align}
where $E({\rm M}, p)$ is the minimum energy of the compound in prototype $p$ per number of sites $n$ in the metal sub-lattice, i.e. the number of TMs in the unit cell.
The offset energy for each TMD, $E_\mathrm{GS}({\rm M})$, is the minimum energy across the prototype space $E_\mathrm{GS}({\rm M}) = \min_p E({\rm M}, p)$.
As the starting point are ML geometries, the present analysis is especially relevant for experimental techniques able to bias the synthesis towards atomically thin films~\cite{Wang2021AtomicOutlook}.
At the same time, the results presented here for monolayer could, in principle, be extrapolated to bulk layered TMDs, as the binding energy between the layers (typically around $\SI{10}{meV/atom}$ for TMDs~\cite{Irving2017a,Levita2014a}) does not usually affect the single layer phase behaviour~\cite{Silva2020a}.

Finally, the lattice stability definition in \cref{eq:MS2_formen} is easily adapted to fixed metal \ce{M} and varying chalcogenide \ce{X}, i.e. $E_\mathrm{F}(\ce{X}, p)$ normalised to the number of sites in the chalcogenide sub-lattice.

\begin{figure*}[!ht]
    \centering
    \includegraphics[width=\textwidth]{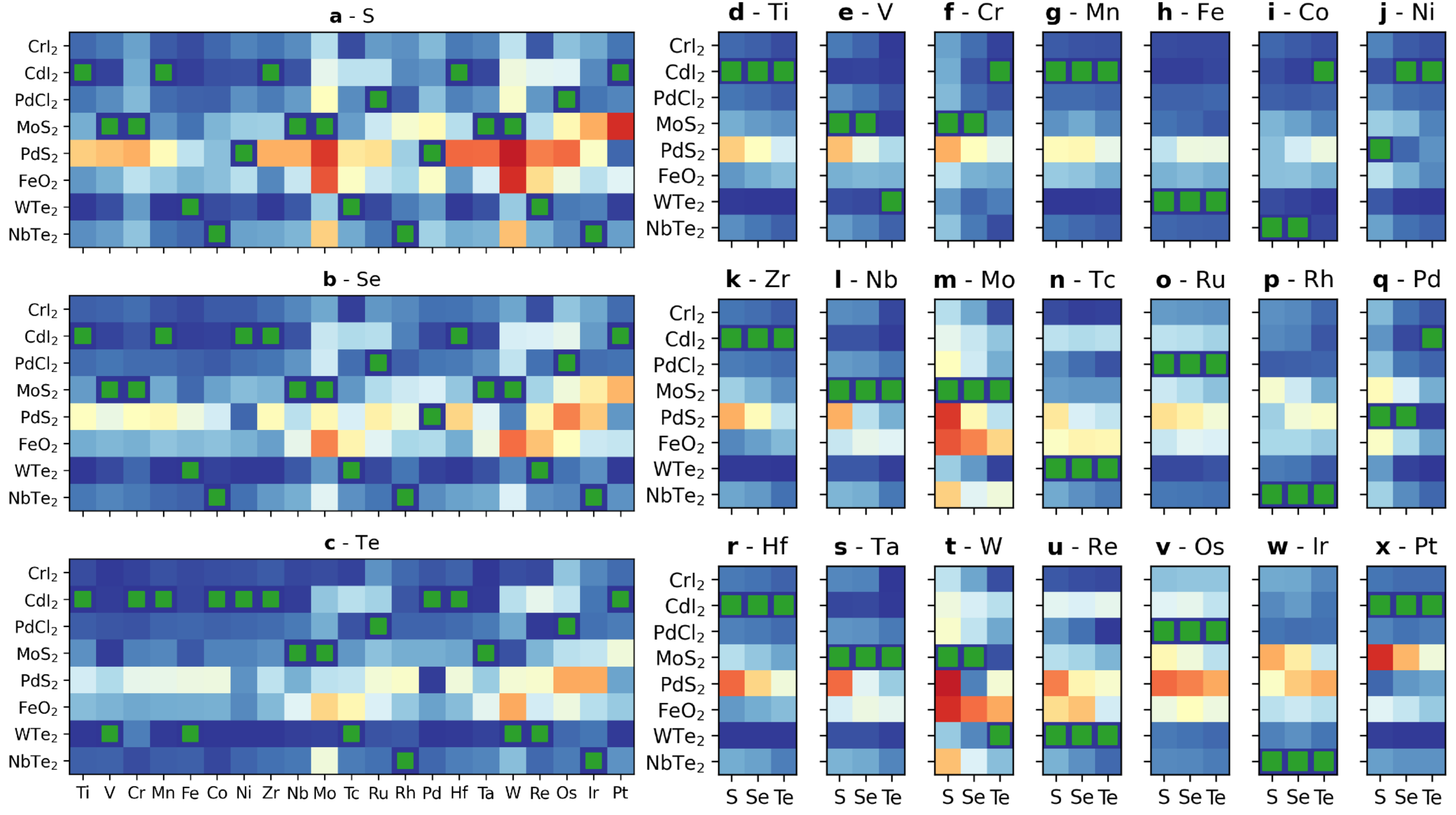}
    \caption[Lattice stability of $MX_2$ compound in the prototypes]{
    Lattice stability of $MX_2$ compounds for (a) $X=\mathrm{S}$, (b) $X=\mathrm{Se}$, and (c) $X=\mathrm{Te}$ and (d-x) for fix chalcogenide and fix TM.
    The prototypes (shown in \cref{fig:p_table-TM_calc}a-h) are indicated on the $y$ axis.
    The varying element ($M$ or $X$) is indicated on the $x$ axis.
    The color scale reports the energy above the ground state in eV per lattice sites from deep blue ($E_\mathrm{F}=\SI{0}{eV/site}$) to bright red ($E_\mathrm{F}=\SI{2}{eV/site}$).
    Green squares mark GS within the eight prototypes, defined by $E_\mathrm{F} = 0$ in the corresponding column.
    The numerical value associated with each entry in panels (a-c) and (d-x) are reported in the SI Table SIV and Table SV, respectively.
    }
    \label{fig:stab_matr}
\end{figure*}

\Cref{fig:stab_matr}a-c report the energy above the ground state per lattice site defined in \cref{eq:MS2_formen} for the selected TMs at fixed chalcogenide and \cref{fig:stab_matr}d-x for the selected chalcogenides at fixed TM.
Each column shows the energy above the ground state of the given TMD \ce{MX_2} in the eight prototypes with respect to the identified 2D GS (green squares). Blue shades designate low energy prototypes, while yellow to red shades designated high energy prototypes.

Known coordination trends in layered compounds are identified correctly. TMDs based on $d^2$-metals (Ti, Zr and Hf) favour the octahedral coordination of the p-\ce{CdI2} prototype (\cref{fig:p_table-TM_calc}b) for all chalcogenides~\cite{Jain2013a}.
The GS of transition metal sulphides based on $d^4$ metals (Cr, Mo, W) is the prismatic prototype p-\ce{MoS2} (\cref{fig:p_table-TM_calc}a), while the GS coordination switches to octahedral for \ce{CrTe2} and \ce{WTe2} (see \cref{fig:stab_matr},t)~\cite{Ong2008,Jain2013a}.

An important chemical trend emerges by comparing the lattice stability in the different chalcogenide spaces in \cref{fig:stab_matr}a-c: the average energy above the GS reduces from the sulphides to the tellurides.
This trend can be understood in terms of the evolution of the bond character between the metal and chalcogenide: the bonds in tellurides are more covalent than in sulphides.
The charge redistribution in these strongly covalent bonds can change the GS prototype, e.g. by enhanced metal-metal bonding~\cite{Mar92WTe2} or by reduced energy penalties of non-GS coordination environments.

\Cref{fig:stab_matr}d-x reports the lattice stability for fixed metal and varying chalcogenides in all considered prototypes.
Trends for varying anion \ce{X} are simple compared with the varying metal case: in most cases, the same ground state is found for S-, Se- and Te-based TMDs and the GS prototype follows the metal period, e.g. all $d^2$ metals (Ti, Zr, Hf) favour p-\ce{CdI2} for any chalcogenide.
The origin of this regular behaviour can be explained in terms of coordination chemistry: the prototype stability is mostly dictated by the $d$ manifold of the metal. This has implications for alloying on the chalcogenide site. In the majority of cases, where the GS geometry is the same for two chalcogenides, alloying on the \ce{X} site at fixed metal should be thermodynamically favourable to tailor properties, as discussed further below.
On the other hand, those rarer cases where the GS prototype changes with the chalcogenide, e.g. the W-based TMDs in \cref{fig:stab_matr}t, could harbour interesting polymorphism and phase transitions as a function of the concentration of the substituting element; this case is discussed in detail in \cref{sec:WSeTe_alloy} and compared with  experimental data.

Finally, it is important to realise the scope of validity and  possible sources of errors in the dataset presented here. 
Spin-polarised DFT calculations are used. Hence, non-magnetic and ferromagnetic GS are correctly described.
Antiferromagnetic (AFM) orderings are not considered, as calculations are performed in cells comprising a single TM site.
To the best of the authors knowledge, the only AFM orderings for the considered stoichiometry are reported for \ce{NiS2} and \ce{MnS2}~\cite{Yu2015thermoChem}.
While important for materials properties, AFM GS in layered TMDs are usually almost degenerate in energy with FM states~\cite{Yu2015thermoChem} and represent a second order effect in phase stability that has been excluded here for the sake of manageable computational effort. 
Moreover, no Hubbard correction (GGA+U) is included here.
The effect of Hubbard U on the relative total energy for the considered TMD stoichiometry is negligible~\cite{Yu2015thermoChem}, but a detailed benchmark must be carried out when applying our protocol to different stoichiometries, as discussed in the Methods section.
Moreover, as the \ce{M-X} bonds develop a more covalent character from $\ce{X}=\ce{S}\rightarrow\ce{Te}$, pronounced charge redistribution may occur in specific orderings of \ce{Q_{1-x}M_xTe_2} systems, yielding a significant change in formation energy, i.e. the formation of ternary compounds. 
This deviation from the pristine compounds behaviour cannot be capture by the metric defined below.
A telluride case where the predictions of our metric are verified is discussed in \cref{sec:WSeTe_alloy}, but care must nonetheless be taken when exploring the tellurides more generally.

%------------------------------------------------------
% Ideal solid solution limit
%------------------------------------------------------
\section{Metastability metric in the ideal solid solution}\label{sec:mstab_metric}
An intuitive approach to explore which metals are likely to mix in a given chalcogenide host (and vice-versa) is the ideal solid solution limit, a non-interacting model based on the lattice stability of pristine, binary TMDs defined in \cref{eq:MS2_formen}.
As for the lattice stability in \cref{eq:MS2_formen}, we focus first on substitution on the TM site; the generalisation to the chalcogenide site is straightforward and briefly outlined afterwards.
Given a binary pseudo-alloy on the metal site in a prototype $p$, \ce{M_x Q_{1-x}X_2}$|_p$, the ideal solid solution represents a model with negligible interactions between the fraction $x$ of sites occupied by \ce{M} and the remaining $1-x$ sites occupied by \ce{Q}.
In the ideal solid solution model, the behaviour of a prototype $p$ in energy-composition space is represented by the line connecting the energy above the ground state of \ce{QX_2} at $x=0$ with the energy above the ground state for \ce{MX_2} at $x=1$ in the same prototype, e.g. the elements ($\ce{Q},p$) and ($\ce{M},p$) of the matrix in \cref{fig:stab_matr}a-c, respectively.
Hence, in the ideal solid solution model, the energy above the ground state of a mixed configuration at concentration $x$ is given by:
\begin{equation}
\label{eq:SS_formen}
    E_{\ce{Q},\ce{M}, p}^0(x) = x E_\mathrm{F}(\ce{M},p) + (1-x) E_\mathrm{F}(\ce{Q},p).
\end{equation}
By construction, this energy is exactly zero everywhere if \ce{M} and \ce{Q} share the same GS structure $p$, i.e. $E_\mathrm{F}(\ce{M},p) = E_\mathrm{F}(\ce{Q},p) = 0$.
In any other case, the energy will be positive: suppose the metal \ce{M} has a GS geometry $p'\neq p$, the fraction $x$ of material \ce{MX_2}$|_p$ would transform into $p'$ to reach equilibrium at zero temperature.

The model effectively quantifies the metastability at zero temperature of alloys in a selected prototype $p$ as a function of concentration $x$.
By construction, this model cannot predict stable mixtures, i.e. negative formation energies, but can be used to estimate the likelihood of solubility and phase separation in a system: the lower the metastability of the solid solution model, the smaller any stabilising mechanisms must be to enable alloy formation under synthesis conditions.
For example, entropy could stabilise solid solutions at finite temperature.
The equilibrium of an alloy in the prototype $p$ at temperature $T$ is determined by the free energy $F_{\ce{Q},\ce{M},p}(x,T) = E^0_{\ce{Q},\ce{M},p}(x) - TS(x)$, where the configurational entropy of an ideal binary alloy is
$S(x) = -k_b[x \log x+(1-x) \log(1-x)]$.
It weights all possible configurations of the two atom types on the metal sub-lattice equally and is a function of the concentration $x$ only, independent of the elemental pairs~\cite{ford2013statistical}.
This stabilization mechanism will be discussed in detail below in relation to experimental synthesis temperatures.
At zero temperature, electronic effects may likewise stabilise orderings, especially in the Te-based TMDs, where covalent bonds may lead to strong mediated interactions between metal sites~\cite{Silva2022pettifor,Mar92WTe2}.

\begin{figure}[!htb]
    \centering
    \includegraphics[width=\columnwidth]{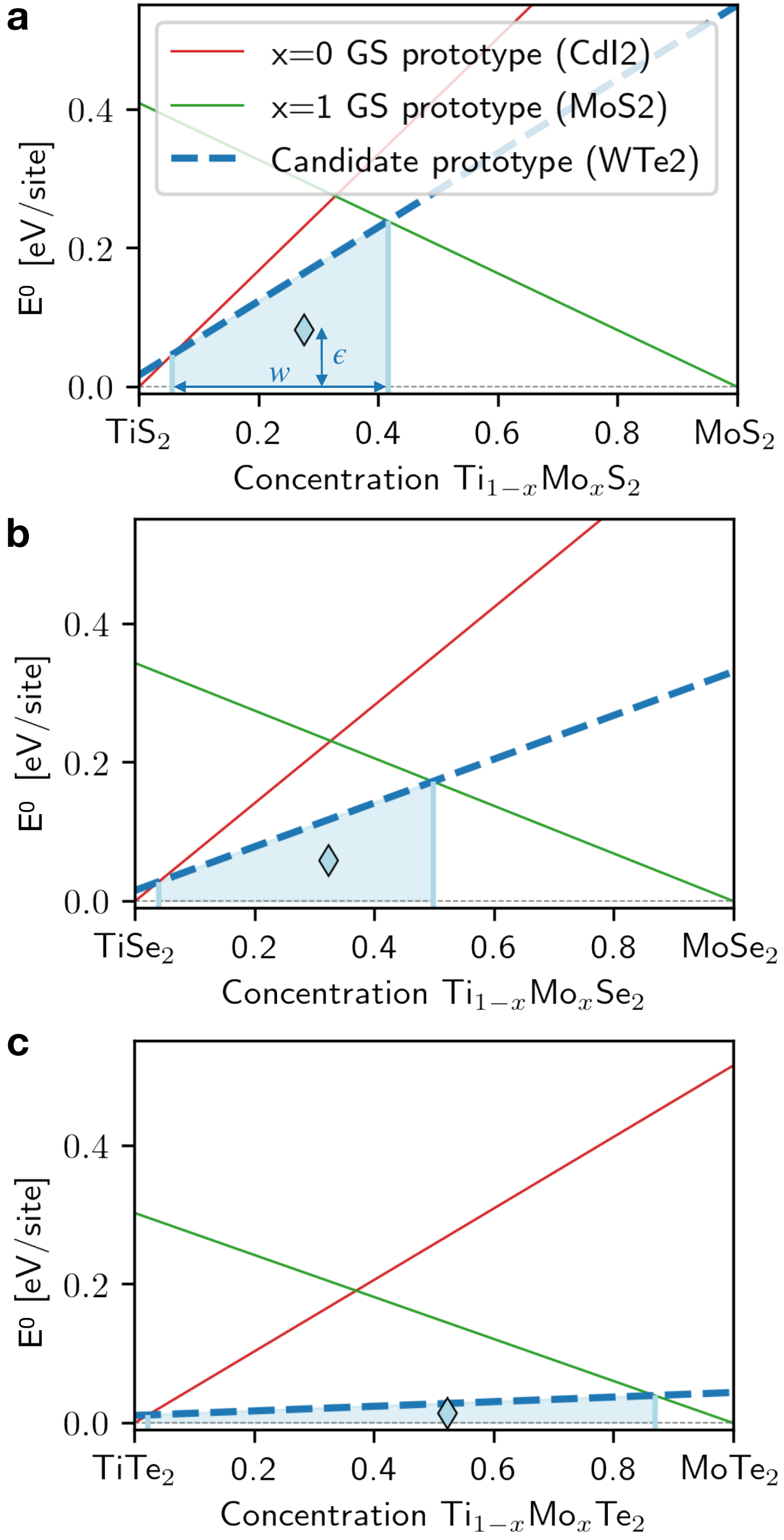}
    \caption[Examples of metastability metric]{
    Construction of the metastability metric for TMD compounds based on Mo and Ti transition metals and different chalcogenides (a) (Ti:Mo)S$_2$ (b) (Ti:Mo)Se$_2$, (c) (Ti:Mo)Te$_2$.
    The ground state prototypes are CdI$_2$ for Ti$X_2$ (solid red line) and MoS$_2$ for Mo$X_2$ (green line).
    The candidate prototype is WTe$_2$ (dashed blue line).
    Light blue areas highlight the extent of the metastability window in the energy above the ground state - concentration $(x, E)$ space.
    Blue diamonds mark the centroids of the metastability window.
    The height of the centroid $\epsilon$ and the window width $w$ are the arguments of the ranking function in \cref{eq:fit_function}.
    }
    \label{fig:metric_example}
\end{figure}

%------------------------------------------------------
%\subsection{Metastability Metric}\label{sec:solub_metric}
A metric in the composition-energy space is used to compare the relative metastability of pseudo-binary alloy candidates.
We focus first on metal site substitutions and consider a prototype $p$ and two chalcogenides \ce{MX_2} and \ce{QX_2} with GS prototype 
%$p_\ce{M}$  % !! Generates error with _ and single letter.
$p_\mathrm{M}$ % Use normal mathrm
and $p_\mathrm{Q}$, respectively.
The convex hull across all phases in the concentration-energy space is the line $E=0$ connecting the energies of the end-members in their respective GS prototypes, i.e. the dashed gray lines in \cref{fig:metric_example}.
A point on this line at the fractional concentration $x \neq 0,1$ represents a phase separating system where the fraction $x$ of \ce{MX_2} is in its GS prototype $p_\mathrm{M}$ and the remaining $1-x$ is in its own GS $p_\mathrm{Q}$.
For a configuration to be stable, its energy must be lower than this hull.
As our model by definition cannot break this hull, we characterise the metastability of a model alloy by its positive energy above the ground state, i.e. its distance from the hull~\cite{Sun2016a}.

We define a descriptor intended to capture the energetic ``disadvantage'' of a particular prototype $(p, \ce{Q}, \ce{M})$ relative to the relevant binary ground states as follows.
The metastability window of the $(p, \ce{Q}, \ce{M})$ triplet is defined as the range of concentration $x$ where the distance from the hull given by \cref{eq:SS_formen} within the prototype $p$ is lower or equal to the distance from the hull within the GS prototypes $p_\mathrm{M}$ and $p_\mathrm{Q}$, as shown by blue regions in \cref{fig:metric_example}.
The metastability metric characterises this window in term of its width $w$ along the concentration axis (see light-blue vertical lines in \cref{fig:metric_example}) and the height of the energy penalty centroid of the window (see light-blue diamond in \cref{fig:metric_example}).
The same construction applies to substitution on the chalcogenide site at fixed metal, i.e. two compounds \ce{MX_2} and \ce{MY_2} with GS prototype $p_\mathrm{X}$ and $p_\mathrm{Y}$, respectively.

Let us apply this construction to an example: consider the solid solution model of the (Mo:Ti)S$_2$ alloy shown in \cref{fig:metric_example}a. 
The solid red line refers to the energy distance from the hull along the tie line \ce{Ti_{1-x}Mo_xS_2} of the p-\ce{CdI_2} prototype, which is the GS of \ce{TiS_2} at $x=0$, i.e. $E_{\ce{Ti}, \ce{Mo}, \ce{CdI}_2}^0(x=0) = 0$ in \cref{eq:SS_formen}.
The solid green lines refers to the ground state of \ce{MoS2}, with $E_{\ce{Ti}, \ce{Mo}, \ce{MoS}_2}^0(1) = 0$.
The dashed blue line refers to the candidate prototype p-\ce{WTe_2}, which is the GS of neither, i.e. $E_{\ce{Ti}, \ce{Mo}, \ce{WTe_2}}^0(x) \neq 0, \forall x$.
The distance from the hull of these prototypes varies as a function of the concentration: the GS prototypes are favoured near the respective end-members, e.g. p-\ce{MoS2} in the range $x \in [0.4,1]$ in \cref{fig:metric_example}a.
The candidate prototype p-\ce{WTe_2} provides a lower energy metastable solution than the two end member GS prototypes in the range $x \in [0.1,0.4]$: the corresponding metastability window is assigned the width $w$ and the energy penalty $\epsilon$ highlighted in \cref{fig:metric_example}a.
The metastability metric is sensitive to the chemistry of the system also at fixed cation: the metric evolves for different cations $\ce{X}=(\ce{S},\ce{Se},\ce{Te})$ as shown in \cref{fig:metric_example}a-c.

The possible scenarios are the following:
(i) When the two TMDs share the same prototype GS, the distance from the hull in that prototype is zero everywhere and the metastability window extents from $x=0$ to $x=1$.
In this case, solubility is likely and the metastability metric is $w=1, \epsilon=0$.
(ii) When the candidate prototype $p$ is the GS for one of the pristine compounds, the metastability window extends from the extremal concentration, $x=0$ or $x=1$, up to the intercept with the ground state of the other compound.
(iii) For non-GS prototypes, there could be a metastability window of finite width $0<w<1$ and energy penalty $\epsilon>0$, or the metastability window does not exist, when the distance from the hull of the candidate prototype is higher than either GS prototypes for any concentration.
In the latter, phase separation in that prototype is likely and the metastability metric is $w=0, \epsilon>0$.

Applying the construction depicted in \cref{fig:metric_example} to all TM pairs yields a $N_{\ce{M}} \times N_{\ce{M}}$ matrix, for each prototype $p$ and each chalcogenides \ce{X}.
Conversely, applying the construction to all chalcogenide yields a $N_{\ce{X}} \times N_{\ce{X}}$ matrix, for each prototype $p$ and each of the $N_{\ce{M}}=21$ metals.
Each entry of these \textit{metastability matrices} is a $2 \times 2$ matrix containing the bounds of the metastability window and the energy above the ground state in \cref{eq:SS_formen} evaluated at the metastability limits, i.e. minimum and maximum hull-distance within the window.
The matrices associated with each prototype are reported in Section III of the SI.

%------------------------------------------------------
\section{Optimal Prototypes for alloys}\label{sec:optimal}
Given two TMDs, we identify the prototype most receptive for substitutional alloying on the metal or chalcogenide site by ranking the metastability metric of all TM$_1$-TM$_2$-prototype (or X$_1$-X$_2$-prototype) triplets.
The following parametric function assigns a single value to the metastability windows
\begin{equation}
\label{eq:fit_function}
   \Gamma_\zeta(w,\epsilon) = \zeta^2 \frac{\sqrt{w}}{\zeta^2+\epsilon^2
   },
\end{equation}
where $w$ is the width of the metastability window and the energy penalty $\epsilon$ is the hull-distance of the centroid defined by the window in the energy-concentration space, i.e. blue diamonds in \cref{fig:metric_example}.
The ranking function is normalised between zero and one, $\Gamma_\zeta(w,\epsilon): w \in [0,1] \,\epsilon \in [0,\infty] \to [0,1]$: it associates zero to "bad" candidates and one to "good" candidates.
In detail, all zero-width windows are mapped to zero, $\Gamma_\zeta(0,\epsilon) = 0 ~ \forall\epsilon$, while the highest score is assigned to the combination of maximum width and null energy penalty, i.e. $\Gamma_\zeta(1, 0) = 1$.
Effectively, the function encourages large metastability windows $w$ and discourages large energy penalties $\epsilon$.
Details regarding the ranking function and the selection of the appropriate weight, $\zeta=\SI{0.080}{eV/site}$ for the present dataset, are reported in Section V of the SI.

\begin{figure*}[!htb]
    \centering
    \includegraphics[width=\textwidth]{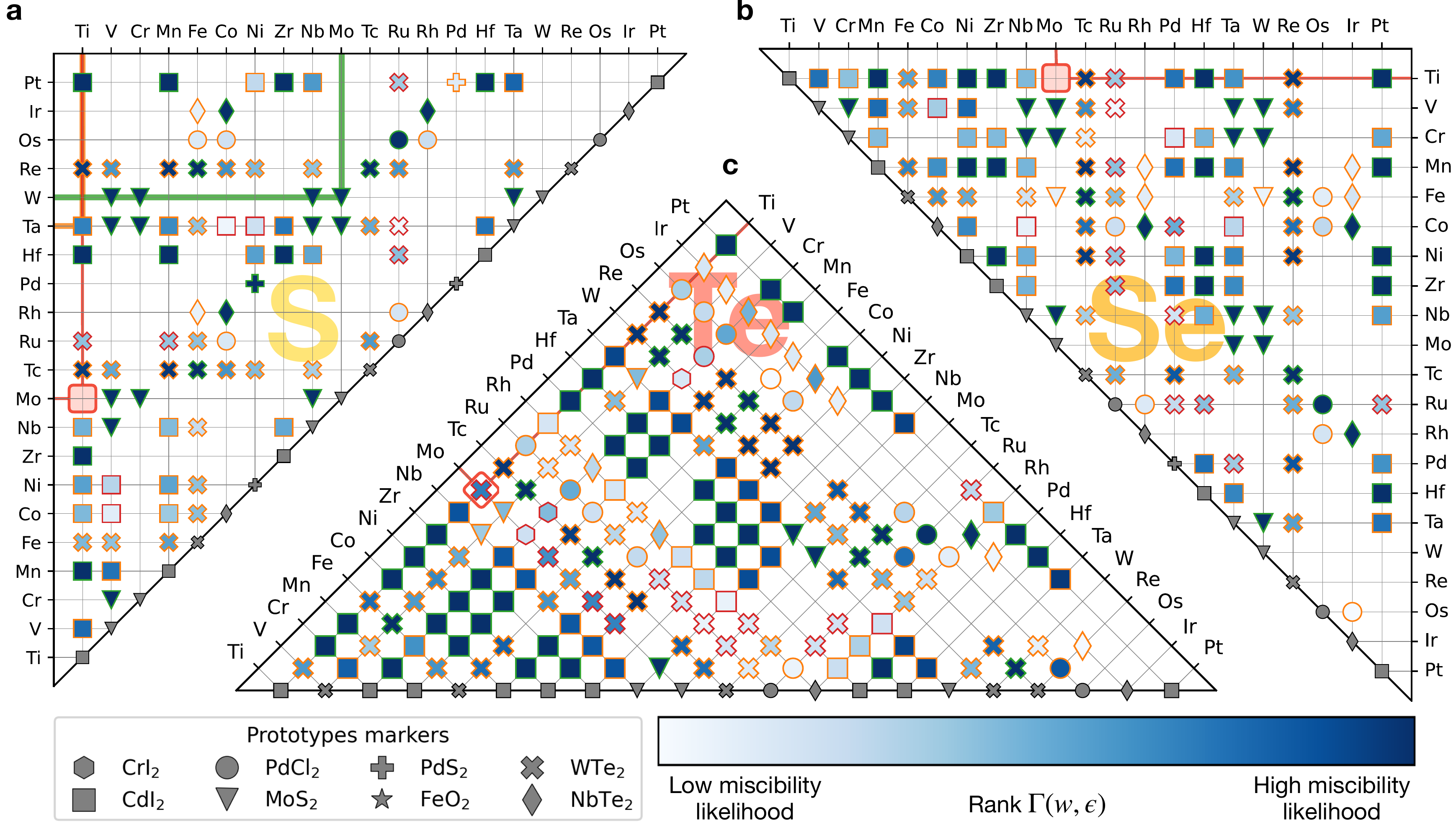}
    \caption[Optimal prototype host for TM pairs]{
    Pettifor maps for the optimal prototype for $Q_xM_{1-x}X_2$ pseudo-binary alloys at fixed chalcogenide (a) $X=\mathrm{S}$, (b) $X=\mathrm{Se}$, and (c) $X=\mathrm{Te}$.
    The prototypes are ranked using $\Gamma_\zeta$ with $\zeta = \SI{0.080}{eV/site}$.
    The color-code refers to the likelihood of the alloy according to the ranking in \cref{eq:fit_function}, see the colorbar on the bottom right.
    The edge color of each marker indicates whether the optimal prototype is the ground state of both (green), one (orange) or neither (red) the pristine TMDs comprising the $(M:Q)X_2$ mixture.
    Marker-prototype correspondence is reported in the legend at the bottom left.
    Gray markers on each diagonal of each panel a,b, and c report the GS prototype of the corresponding pristine TMD $MX_2$.
    Green, orange and red lines serve as a guide to the eye toward the entries corresponding to the examples discussed in the main text.
    For a version without rotation and example lines see SI Section VI Figure S13-16.
    }
    \label{fig:optimal_proto}
\end{figure*}
The optimal prototypes for substitution on the metal site are shown in \cref{fig:optimal_proto} for each pair of transition metals.
The colour code of each entry shows the ranking of \ce{Q_{1-x}M_xX_2} in the optimal prototype, see markers legend.
Additionally, the edge of each marker indicates whether that prototype is the ground state of both (green edge), one (orange edge), or neither (red edge) pristine compounds.
\Cref{fig:optimal_proto} provides a visual tool to navigate the possible mixtures of transition metals within the chalcogenide planes.
Large blue marks in \cref{fig:optimal_proto} indicate a high rank (small energy penalty and wide metastable window) and, thus, that miscibility between the two metals within the chalcogenide host is likely.
On the other hand, white marks indicate a low score (high energy penalty and small metastable window) that likely results in miscibility gaps.

The distinction between likely-mixing and likely-separating systems can be further constrained by extending the Hume-Rothery rules to our case~\cite{Abbott,Silva2022pettifor}: miscibility between transition metals within a chalcogenide host is expected if the lattice mismatch between the pristine compounds is less than 15 \% ~\cite{Abbott} (see SI Section IV for definition and values of the mismatch in these compounds) and the energy above the ground state of the optimal prototypes is below a threshold of $E = \SI{120}{meV/site}$, as metastable compounds within this range have been observed experimentally~\cite{Sun2016a}. 
As a result, \cref{fig:optimal_proto} features “missing elements” where the optimal prototypes are unlikely to be receptive to alloying due to large lattice mismatch or high energy above the ground state. A different layout of the optimal prototype maps, with the full information on the energy penalty and window size, is reported in SI Figure S16-17.
Note that the maps become more populated going from sulphides to tellurides.
This is in agreement with the lowering of the energy landscape with increasing covalency that is also seen in \cref{fig:stab_matr}.
For a quantitative visualisation of this trend see Figure S18 in the SI.

As an example of how to navigate the map, consider the pseudo-binary \ce{Mo_{1-x}W_xS_2}.
Following the green lines in the sulphides map, \cref{fig:optimal_proto}a, leads to a deep blue triangle with green edges, indicating the maximum ranking for p-\ce{MoS2}, which is the GS of both compounds.
This corresponds to the maximum likelihood to mix.

As another example in the sulphides, consider the \ce{Ti_{1-x}Ta_xS_2} pseudo-binary, whose entry is highlighted by orange lines in \cref{fig:optimal_proto}a.
The map reports as the optimal geometry p-\ce{CdI_2}, which is the GS of \ce{TiS_2}, but not of \ce{TaS_2} (GS prototype p-\ce{MoS_2}); hence the orange edge.
The marker color is blue (but lighter than in the best-rank previous example Mo$_{1-x}$W$_x$S$_2$), signalling that alloying is still likely even in the non-native host.
This prediction is discussed in detail in the next section.

Finally, Mo-Ti-based TMD alloys provide an example of varying phase behaviour in different chalcogenide spaces.
The entries in the sulphide, selenide and telluride cases are highlighted by red lines and squares in \cref{fig:optimal_proto}a,b,c.
In the S and Se spaces, the entry is missing, signalling that the metals are likely to phase separate according to the generalised Hume-Rothery rules.
But, the likelihood of forming an alloy increases in the tellurides, as signalled by the light-blue cross (p-\ce{WTe_2}) in \cref{fig:optimal_proto}c.
This trend is consistent with the low lattice stability penalty in tellurides seen in \cref{fig:stab_matr}c, and with the evolution of the metastability metric reported in \cref{fig:metric_example}: the lattice stability of p-\ce{WTe_2} on the Mo-rich side $x=1$ reduces significantly along $\ce{S}\rightarrow\ce{Se}\rightarrow\ce{Te}$ from $E_\mathrm{F}(\ce{Mo}, \ce{WTe_2})=\SI{0.55}{eV/site}$ for S over $\SI{0.33}{eV/site}$ for Se to $\SI{0.04}{eV/site}$ for Te.
Consequently, the centroid energy and metastability window width (light blue diamond and area in \cref{fig:metric_example}) become lower and wider, respectively, yielding a favourable ranking $\Gamma$ in the tellurides. 
The stability of the distorted octahedral structure of \ce{WTe_2} has been attributed to an increase in direct metal-metal bonding~\cite{Mar92WTe2}; we speculate that the same argument could apply for MoTe$_2$ in p-\ce{WTe2}, given the chemical similarity between Mo and W.
This pseudo-binary alloy is further characterised in the next section.

As a first benchmark, the information in \cref{fig:optimal_proto} can be compared with alloys reported in the literature.
We first focus on alloys of the most studied pristine compound in the TMD family: MoS$_2$.
Zhou and coworkers~\cite{Zhou2018} recently reported synthesis of (Nb:Mo)S$_2$ MLs, which is shown as likely to mix in \cref{fig:optimal_proto}a.
However, the same work reports a (Mo:Re)S$_2$ ML alloy, whose metastability window is small and high in energy (see \cref{fig:optimal_proto}a and SI Figure~S2).
Another recent report~\cite{Zhu2019d} characterises (V:Mo)S$_2$ MLs experimentally, which is also a TM pair likely to mix according to our analysis.
The first example mentioned above, (Mo:W)S$_2$ (green lines in \cref{fig:optimal_proto}a), has also been realised in experiments~\cite{Chen2013,Xia2021}.
Finally, V-doped WSe$_2$ and (Mo:W)Se$_2$ alloys have been recently synthesised~\cite{Stolz2022VWSe2_alloy,Ahmad2019MoWSe_alloy,Susarla2017MoWSSe_alloy} as non-\ce{MoS2}-based examples that are both indicated as likely miscible alloys in \cref{fig:optimal_proto}b.

\begin{figure*}[!htb]
    \centering
    \includegraphics[width=\textwidth]{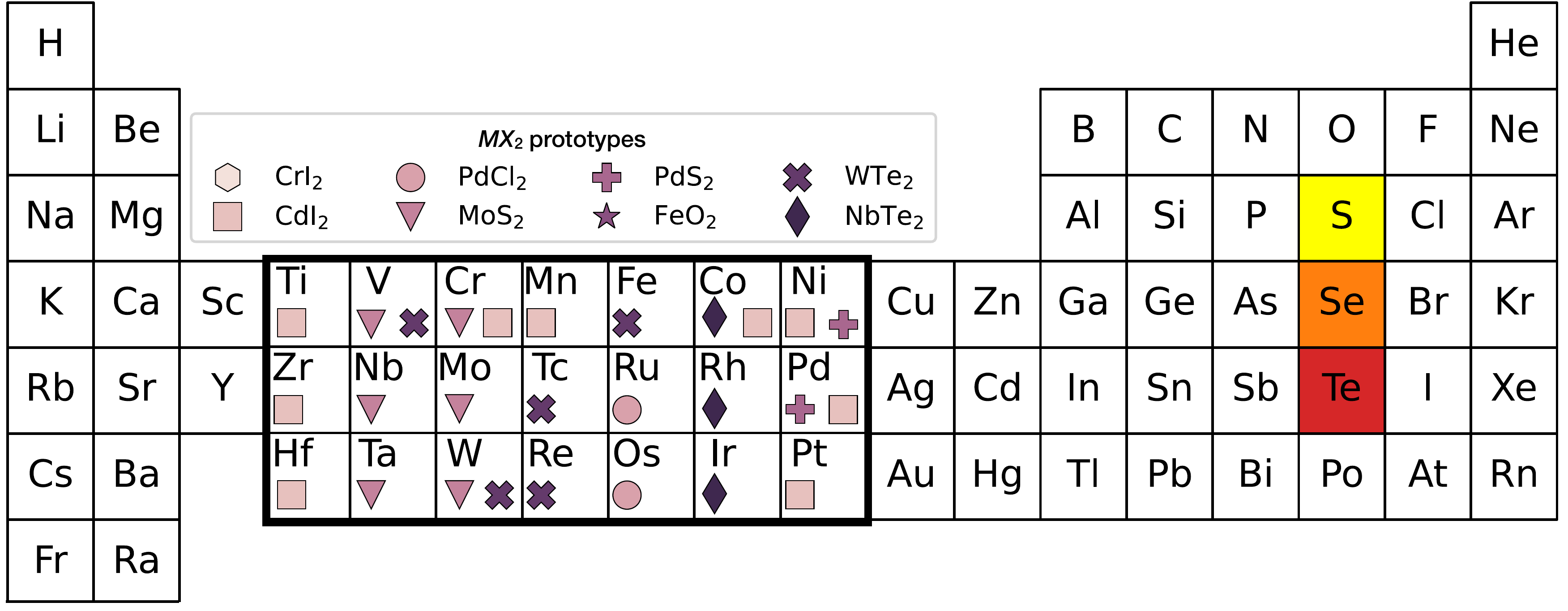}
    \caption[Polymorphism with varying chalcogenide at fixed TM]{
    Polymorphism for $M(X_xY_{1-x})_2$ pseudo-binary alloys at fixed metal $M$ (entries highlighted by the thick black rectangle in the periodic table) and varying chalcogenide $X,Y$ (colored entries in the periodic table).
    Marker-prototype correspondence is reported in the legend at the top center.
    The leftmost symbol in each metal $M$ entry correspond to the most receptive prototype for any substitutional alloy on the chalcogenide site, i.e. $M(\mathrm{S}_x\mathrm{Se}_{1-x})_2$, $M(\mathrm{S}_x\mathrm{Te}_{1-x})_2$, and $M(\mathrm{Se}_x\mathrm{Te}_{1-x})_2$.
    The rightmost marker (if any) correspond to the second best prototype for chalcogenide mixtures.
    For the optimal prototype matrices underlying this polymorphism map see the SI Figure S17.
    }
    \label{fig:chalc_polimorph}
\end{figure*}

The optimal prototype is not very sensitive to a change of the chalcogen atom at fixed metal, as discussed in \cref{sec:chem_space}.
The optimal geometry for pseudo-binary alloys on the chalcogenide site is, therefore, predominantly the common GS prototype.
This allows to report the most likely prototypes for a given TM across the chalcogenide spaces in \cref{fig:chalc_polimorph}, which is a condensed version of the maps in \cref{fig:optimal_proto}. The most and second-most receptive prototypes for alloys \ce{M(X_xY_{1-x})2} are given for each of the considered cations. \cref{fig:chalc_polimorph} is convenient to gauge likely TM coordinations for each of the transition metals.

Consider the Ti entry as an example: the GS is p-\ce{CdI_2} for S, Se and Te (see \cref{fig:stab_matr}d) and thus alloying on the chalcogenide site is most likely to occur in this prototype.
When GS prototypes differ between chalcogenides at fixed metal, alloy possibilities in non-native prototypes may arise. 
Tungsten exhibits this type of polymorphism: the dominant prototype (left symbol) is p-\ce{MoS_2}, the GS prototype of \ce{WS_2} and \ce{WSe_2}; but a second symbol is added on the right for the homonymous prototype p-\ce{WTe_2}.
\cref{fig:chalc_polimorph} can be benchmarked against limited experimental data. The same-prototype alloys Mo(S:Se)$_2$, Mo(S:Te)$_2$ and W(S:Se)$_2$ have been synthesised~\cite{Su2014MoSSe_alloy,Susarla2017MoWSSe_alloy,Liu2021MoTe,Tang2021MoSSe_alloy,Kim2021TMD_alloy}. 
W(Se:Te)$_2$ is a confirmed case of polymorphism between two prototypes. This system is analysed in detail and compared with available experimental data~\cite{Yu2016WSe-Te2_phaseT} in the next section.

%------------------------------------------------------
% TM Ordering
%------------------------------------------------------
\section{Orderings in pseudo-binary alloys}
The phase behaviour predicted by the Pettifor maps in \cref{fig:optimal_proto} is benchmarked by sampling the configurational space at varying concentration with electronic-structure calculations.
For substitution on the metal sub-lattice, the formation energy of a pseudo-binary alloy \ce{M_xQ_{1-x}X_2} is obtained by taking the GS end members as reference for the ordered configuration $\sigma(x)$ at concentration $x$:
\begin{align}
    E_{\ce{Q}, \ce{M}, p}(\sigma(x)) =& \left.E(\sigma(x))\right|_p  \nonumber \\  
          &- x E(\ce{M}, p_{\ce{M}}) \nonumber \\  
          &- (1-x) E(\ce{Q}, p_{\ce{Q}}), \label{eq:alloy_formen}
\end{align}
where $\left.E(\sigma(x))\right|_p$ is the total energy per lattice site of the configuration $\sigma(x)$ in the host lattice defined by the prototype $p$.
$E(\ce{M}, p_{\ce{M}})$ and $E(\ce{Q}, p_{\ce{Q}})$ are the total energies per lattice site of \ce{MX_2} and \ce{QX_2} in their GS prototypes $p_{\ce{M}}$ and $p_{\ce{Q}}$, respectively.
This chemical reference assures that the formation energy in \cref{eq:alloy_formen} at end-member concentration $x=0$ and $x=1$ corresponds to the energy above the ground state reported in \cref{fig:stab_matr}.

The set of geometrically distinct orderings is generated using CASM~\cite{VanderVen2010,Puchala2013,Thomas2013}.
The geometries are fully relaxed, including cell shape and volume.
The relaxation of the cell is needed to accommodate possible lattice mismatch.
This extra degree of freedom, however, may result in non-GS prototypes transforming into more stable ones of similar symmetry~\cite{Thomas2021ComparingGeometry}.
For details see the Methods section.

The following section reports a computational test of the predictions discussed in the previous section (highlighted by coloured lines in \cref{fig:optimal_proto}).
These cases represent one strongly and one weakly phase-separating substitutional alloy on the TM site, an alloy on the TM site with finite-miscibility in a non-native prototype already at zero temperature and a case of polymorphism for alloying on the chalcogenide site.

\begin{figure*}[!htb]
    \centering
    \includegraphics[width=0.9\textwidth]{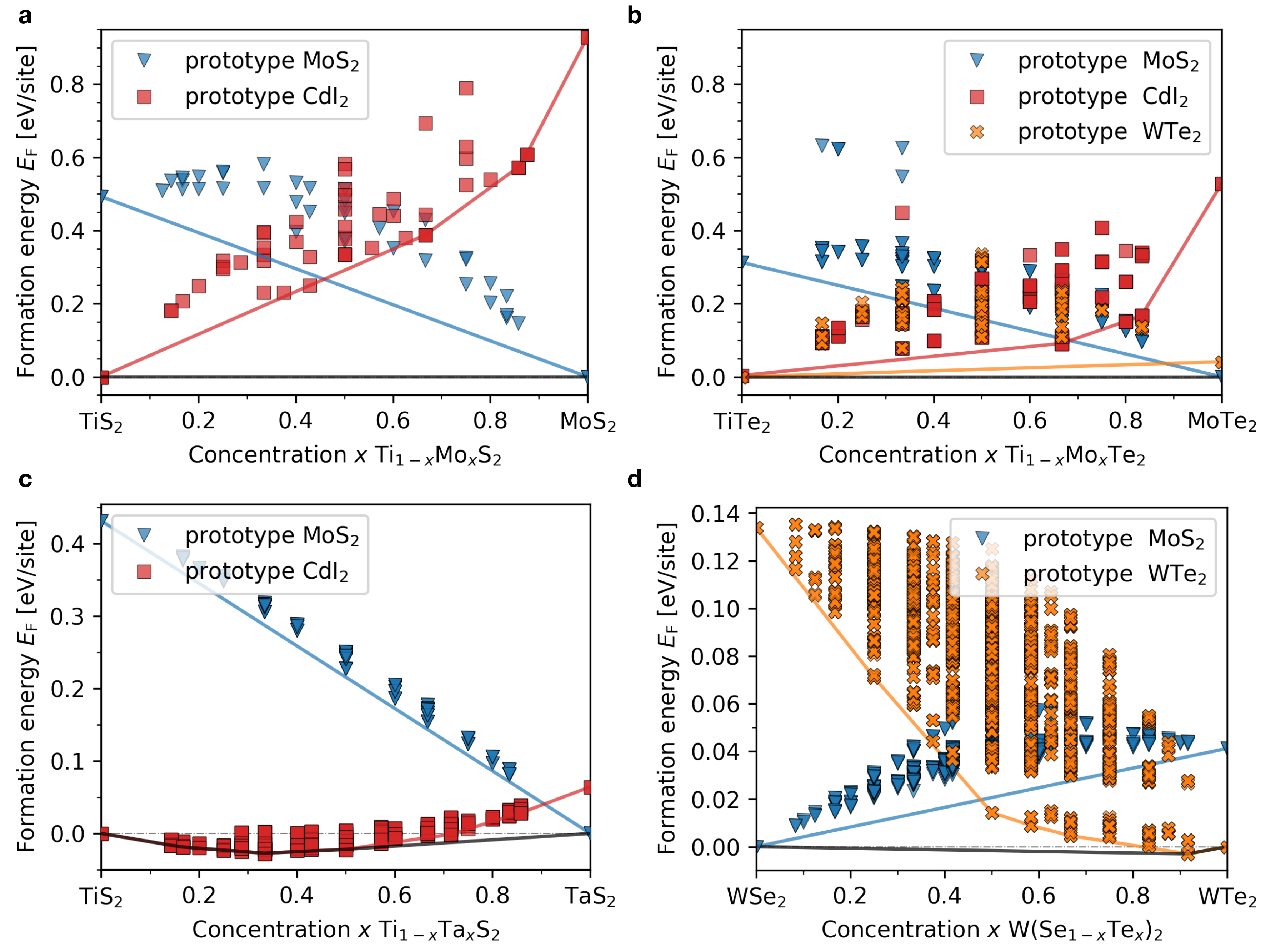}
    \caption[Formation energy of selected binary alloys]{
    Formation energies in eV/lattice site computed from DFT calculation across the tie-line concentrations of the pseudo-binary alloys: (a) (Ti:Mo)S$_2$, (b) (Ti:Mo)Te$_2$, (c) (Ti:Ta)S$_2$, and (d) W(Se:Te)$_2$.
    Different shapes and colors refer to different prototypes as reported in the legend.
    Color-matching solid lines report the convex hull construction within each host, marking the thermodynamic stability at fixed host.
    The black solid line in each plot shows the inter-host convex hull.
    }
    \label{fig:CE_example}
\end{figure*}

\subsection{Strong Phase separating: (Mo:Ti)S$_2$ Pseudo-binary Alloys}
As already discussed, the high lattice stability of Mo in p-\ce{CdI2} and Ti in p-\ce{MoS2}(see \cref{fig:metric_example}a) results in a low ranking of the metastability metric; hence the corresponding missing entry in \cref{fig:optimal_proto}a (or the high-energy solutions in Figure S1 in SI).
The phase separation prediction is confirmed by total energy DFT calculations of the ordered configurations as shown in~\cref{fig:CE_example}a.
No configurations in the p-\ce{MoS_2} prototype (blue symbols) display lower formation energy than the solid solution model (solid blue line).
Within the p-\ce{CdI2} prototype, some configurations display a lower energy compared to the solid solution model, see points on the solid red line.
This electronic stabilisation mechanism, however, is not enough to break the inter-prototype convex hull (dash-dotted gray line at $E=0$), resulting in an overall phase separating system.
The origin of this zero-temperature phase behaviour lies in the different local environment favoured by each TM, as explained in terms of crystal field levels in Ref.~\cite{Silva2020a}.
The effect of temperature is explored in Ref.~\cite{Silva2020a} by means of Monte-Carlo simulations based on a cluster expansion Hamiltonian, trained on the DFT dataset~\cite{VandeWalle2002b}.
The finite temperature phase diagram indicates that the miscibility gap in \cref{fig:CE_example}a closes above the melting temperature of the compounds and that only a small percentage of doping near the end-members is possible due to configurational entropy, in agreement with experimental estimations~\cite{Hsu2001}.

\subsection{Weakly Phase separating: (Mo:Ti)Te$_2$ Pseudo-binary}
While Mo and Ti phase separate within the S host, the metastability metric suggests that alloying should be possible within the Te host, in the p-\ce{WTe_2} prototype (see \cref{fig:optimal_proto}c).
\Cref{fig:CE_example}b reports the benchmark of this prediction.
The formation energy of all configurations are significantly lower in this case, although not enough to break the inter-host hull at zero temperature (black line in \Cref{fig:CE_example}b).
At the same time, this DFT search confirms the higher likelihood of alloying in this case with the miscibility gap expected to close at lower temperatures than reported for (Mo:Ti)S$_2$~\cite{Silva2020a}.

Note that some configurations in the p-\ce{CdI_2} and p-\ce{WTe_2} prototypes overlap on the Ti-rich side of \cref{fig:CE_example}b.
The p-\ce{WTe_2} host can transform into to GS p-\ce{CdI_2} prototype in our computational protocol. Although symmetry is conserved during relaxation, a varying number of point group operations are removed by TM substitutions, which gives sufficient degrees of freedom to change prototype.
The similarity between the two structures driving this transition is quantified in terms of a structural descriptor similar to the SOAP kernel~\cite{Bartok2013SOAP,Bartok2016SOAP} in the SI Section VII.A and Figure~S20. 
Because of this overlap, the convex hull in \cref{fig:CE_example}b can be expressed in terms of the p-\ce{MoS2} prototype and an octahedral-like prototype, comprising the structures derived from p-\ce{CdI_2} and p-\ce{WTe_2}.
The intra-host convex hull of this hybrid prototype (see purple line in SI Figure S21a) lies less than 50 meV from the inter-host convex hull (black line in \cref{fig:CE_example}b) and miscibility should occur up to $x=0.23$ at synthesis temperatures of around $\SI{900}{K}$, see SI section VII.A and Figure S21b for details. 

\subsection{Cross-host miscibility: (Ti:Ta)S$_2$ Pseudo-binary Alloys}
\label{sec:TiTa}
We now test the prediction from the ranking map in \cref{fig:optimal_proto}a presented in \cref{sec:optimal} for inter-host high-miscibility against actual alloy configurations from DFT.
\cref{fig:CE_example}c reports the formation energy of (Ti:Ta)S$_2$ alloys in the p-\ce{CdI_2} (red symbols) and p-\ce{MoS_2} prototypes (blue symbols).
As predicted by the metastability metric, TiS$_2$ and TaS$_2$ segregate in p-\ce{MoS_2}: no configuration lies below the solid solution limit (straight blue line);
see SI Figure S1 for the relative entry in the metastability matrix.
In the p-\ce{CdI_2} prototype, native host for TiS$_2$ but not for TaS$_2$, the alloyed configurations lie below the cross-host solid-solution hull (dash-dotted gray horizontal line) from $x \approx 0$ up to $x \approx 0.7$.
While at zero temperature only the configurations on the inter-host convex hull (black solid line) are stable, the energy scale is small compared to room temperature, suggesting that synthesis of solid-solution alloys in the p-\ce{CdI_2} prototype is achievable experimentally, e.g. with CVD techniques.
Indeed, there are reports of (Ti:Ta)S$_2$ solid solution alloys in the literature ~\cite{THOMPSON1972}, although no crystallography data or solubility limits are available to date.
This experimental confirmation further validates the predictive power of our approach.

\subsection{Polymorphism chalcogenide alloys: W(Se:Te)$_2$ Pseudo-binary}\label{sec:WSeTe_alloy}
\begin{figure}[!htb]
    \centering
    \includegraphics[width=\columnwidth]{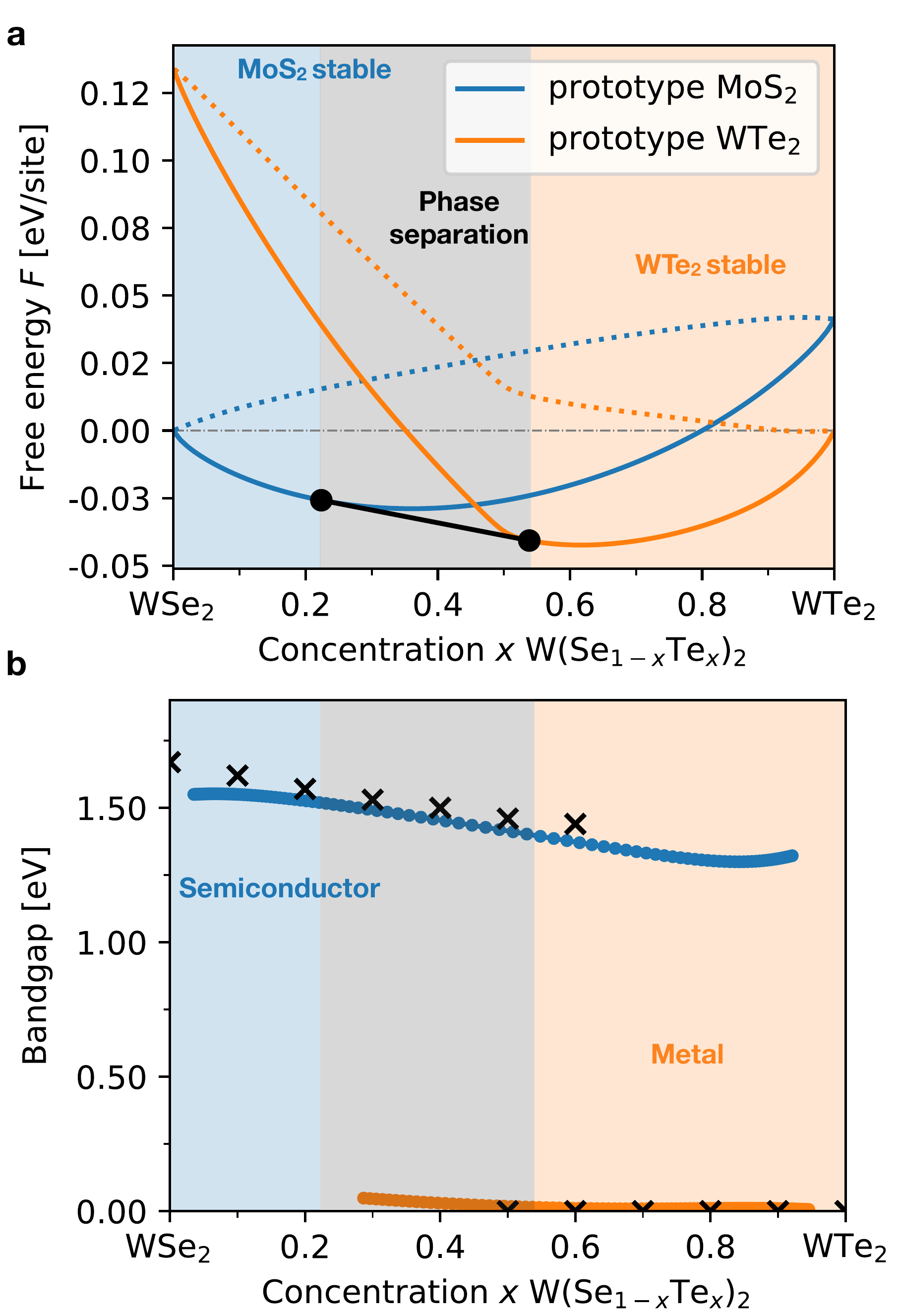}
    \caption[Detail analysis of W(Se:Te)$_2$ pseudo-binary alloy]{
    (a) Thermodynamic stability at finite temperature. The vertical axis reports the free energy $F$ (in eV/lattice sites) at synthesis temperature $T=\SI{900}{K}$\cite{Yu2016WSe-Te2_phaseT} against the concentration $x$.
    The free energy (solid lines) is estimated from the zero temperature convex hull (dotted lines) as outlined in the Methods.
    The solid black line between the dots shows the Maxwell construction defining the phase separating region (gray-shaded area) between the MoS$_2$ (blue-shaded area) and WTe$_2$ stable (orange-shaded area) ones.
    (b) Bandgap as a function of concentration in the MoS$_2$ (blue line) and WTe$_2$ (orange line line) prototypes.
    The shaded area mark the pMoS$_2$-stable, phase-separating, and pWTe$_2$-stable regions defined in (a).
    Black crosses report the experimental values adapted from Ref.~\cite{Yu2016WSe-Te2_phaseT}.
    }
    \label{fig:WTe_props}
\end{figure}

Finally, we discuss in detail an example of alloying on the chalcogen site, which is predicted to show polymorphism.
We focus on W(Se:Te)$_2$, where polymorphism should occur between the GS of WSe$_2$ (p-\ce{MoS_2} prototype) and GS of WTe$_2$ (p-\ce{WTe_2} prototype), as this will allow us to compare directly with experimental data on phase stability and opto-electronic properties.
\Cref{fig:CE_example}d reports the formation energy of ordered configurations, which in this case is given by
\begin{align}
    E_{\mathrm{Se}, \mathrm{Te}, p}(\sigma(x)) =& \left.E(\sigma(x))\right|_p  \nonumber \\  
          &- x E(\mathrm{Te}, \mathrm{WTe}_2) \nonumber \\  
          &- (1-x) E(\mathrm{Se}, \mathrm{MoS}_2). \label{eq:alloy_formen_WTe2}
\end{align}
At zero temperature the system is weakly phase separating in the p-\ce{MoS_2} prototype (blue triangles in \cref{fig:CE_example}d).
There is a small inter-host miscibility window on the Te-rich side in the p-\ce{WTe_2} prototype (orange crosses \cref{fig:CE_example}d): around $x\approx0.9$ an alloy in the p-\ce{WTe_2} prototype is more stable than the phase-separation (see the black line falling below the dash-dotted gray line in \cref{fig:CE_example}d).
In order to compare with experiments, we estimate the finite temperature phase diagram with an approximated Boltzmann sampling, see Methods for details.
\Cref{fig:WTe_props}a reports the free energy of each prototype as a function of concentration at $T=\SI{900}{K}$, compatible with CVD synthesis~\cite{Yu2016WSe-Te2_phaseT}.
Using the Maxwell construction, we estimate miscibility in the p-\ce{MoS_2} prototype up to $x=0.22$ (blue-shaded area in \cref{fig:WTe_props}a), phase separation in the range $x\in[0.22,0.54]$ (gray-shaded area) and miscibility in the p-\ce{WTe_2} prototype from $x=0.54$ (orange-shaded area).
The estimated phase behaviour is in agreement with the experiments in Ref.~\cite{Yu2016WSe-Te2_phaseT}, which reports a phase transition in this pseudo-binary system around $x\approx 0.5$.

The comparison with experiments can be extended to opto-electronic properties.
\Cref{fig:WTe_props}b reports the bandgap in p-\ce{MoS_2} (blue dots) and p-\ce{WTe_2} (orange dots) prototypes as a function of concentration.
See SI section VII.B for details.
The plot is divided in the three phase regions defined in \cref{fig:WTe_props}a.
W(Se:Te)$_2$ in the p-\ce{MoS_2} prototype is a semi-conductor with a bandgap decreasing from 1.55 eV to 1.30 eV as a function of concentration, while W(Se:Te)$_2$ in the p-\ce{WTe_2} prototype is a semi-metal with a vanishingly small bandgap of 0.05 eV at $x\approx 0.5$ that closes for $x>0.6$.
These results are in remarkable agreement with experiments (black crosses in \cref{fig:WTe_props}b): the CVD-grown samples in Ref.~\cite{Yu2016WSe-Te2_phaseT} are semiconductors with bandgaps around 1.5 eV up to $x\approx 0.6$ and turn metallic for higher-concentrations, once the p-\ce{WTe_2} geometry is more stable.

\section{Conclusions}
We presented a systematic analysis of possible substitutional alloys in two-dimensional TMDs on both metal and chalcogenide sites.
The best structural prototypes for alloying are identified via a ranking of a metastability metric.
This ranking, visualised by the chemical space maps shown in \cref{fig:optimal_proto,fig:chalc_polimorph}, provides a guideline for experimental synthesis and an assessment of thermodynamic stability for computational screening of properties of different compounds.

Predictions of phase separating and miscible systems by the metastability metric are in good agreement with experimental reports in the literature and with the systematic computational samplings of ordered structures carried out in this study for selected binary alloys from {\it First Principles}.
While this work focused on TMDs, the methodology developed here can be transferred to any stochiometry and composition, with the caveat that different systems might require a different underlying DFT protocol, e.g. Hubbard corrections for oxides.

The Pettifor maps of optimal prototype in \cref{fig:chalc_polimorph,fig:optimal_proto} can help to identify viable alloy candidates, minimising the trial-and-error attempts and speeding up the discovery of novel materials for nanotechnology.
In particular, these maps could aid CVD synthesis of novel ML alloys in non-native geometries that exhibit desirable properties.

In a wider context, the framework developed here fits in the effort of making chemical intuition quantitative.
The exploration of a large dataset, easily produced with modern DFT methods, allows to rationalise trends across the periodic table and refine known empirical rules or adapt them to new chemical spaces.
Here we showed how the evolution from more ionic bonds in sulphides to more covalent ones in tellurides results in more possibilities for alloying on the metal site.
The analysis at fixed metal and varying chalcogenide confirms the chemical intuition that coordination is dictated by the $d$ manifold of the metal, resulting in the dominance of the same GS prototype for sulphides, selenides and tellurides.
But, our quantitative analysis identifies cases that break this rule and where interesting polymorphism may be found.
These trends are made quantitative by generalised Hume-Rothery rules and the metastability metric, resulting in the compact tool of the Pettifor maps for substitutional alloys in \cref{fig:chalc_polimorph,fig:optimal_proto}.

To summarise, we presented a set of tools and ideas that will hopefully prove a useful guide for computational chemists and experimentalists whilst maping out the under-explored chemical space of two-dimensional TMDs. 

\section*{Methods}
\paragraph{Ab-initio calculations}
The total energy calculations are carried out with the Vienna \textit{Ab Initio} Simulation Package (VASP)~\cite{KresseAv1996,Kresse1993,Kresse1999}, within the PAW framework for pseudo-potentials~\cite{Blochl1994}.
The generalised-gradient-approximation to DFT as parametrised by Perdew, Burke, Ernzerhof~\cite{Perdew1996} was used in this work.
The Kohn-Sham orbitals are expanded in a plane-wave basis with a cutoff of $E_{\mathrm{cutoff}}=\SI{650}{eV}$ and the BZ is sampled with a $17 \times 17 \times 1$ mesh.
The electronic density was computed self-consistently until the variation was below the threshold of $\SI{1e-6}{eV}$.
We perform spin-polarised calculation; the electronic structure can converge to non-magnetic or ferromagnetic states, as we consider only primitive unit-cells in our calculations.
For lattice stability calculations, the position of the ions in the unit cell were relaxed until the residual forces were below the threshold $\SI{1e-2}{eV/\AA}$. 
For configurational sampling calculations the position of the ions and the unit cell were relaxed until the residual forces were below the threshold $\SI{1e-2}{eV/\AA}$.
To ensure no spurious interactions between the periodic images, a vacuum of $\SI{20}{\AA}$ was added along the $c$ axis.

Note that while error cancellation in the stoichiometric analysis carried out here makes the Hubbard U correction not necessary, Ref.~\cite{Yu2015thermoChem} shows that this becomes fundamental in modelling thermochemical reactions involving valance changes, as the reaction enthalpy of most sulphurisation reactions is not correctly described at U=0. 

\paragraph{Approximated Boltzmann sampling}
A Boltzmann weighting of computed configurations
\[
\langle X\rangle \approx \frac{\sum X_i\cdot \exp(-\beta\cdot\Omega_i)}{\sum \exp(-\beta\cdot\Omega_i)}
\]
with $\Omega_i = E_i-x_i\cdot\mu$ was used to estimate ensemble averages. In this approach, $\beta$ is a parameter larger than $1/kT$ to compensate for the over-weighting of high energy configurations implied by sampling over only a small part of the configurational space. The parameter $\beta$ was chosen such that $\langle E\rangle$ closely resembles the convex hulls to reflect the expected small dependence of the internal energy on temperature common for solids. See the SI Section VII for more detail.

%%%%%%%%%%%%%%%%%%%%%%%%%%%%%%%%%%%%%%%%%%%%%%%%%%%%%%%%%%%%%%%%%%%%%
%% The "Acknowledgement" section 
%%%%%%%%%%%%%%%%%%%%%%%%%%%%%%%%%%%%%%%%%%%%%%%%%%%%%%%%%%%%%%%%%%%%%
\section*{Supporting information}
Details on database filtering, chemical and coordination space, ranking function, and configurations sampling. Complete tables of lattice stability, metastability matrices of prototypes, mismatch matrices, and optimal prototype matrices.

\section*{Data and Code Availability Statements}
Pristine compounds data (energies and structures) are included in this published article as supplementary information files as a JSON database.
Lattice stability, lattice mismatch, metastability metric and optimal prototype matrices are provided in NumPy machine format.
Configurational sampling data (energies and structures) are included as an archive.

\begin{acknowledgments}
This project has received funding from the European Union's Horizon2020 research and innovation programme under grant agreement No. 721642: SOLUTION.
The authors acknowledge the use of the IRIDIS High Performance Computing Facility, and associated support services at the University of Southampton, in the completion of this work.
DK and JC acknowledges support form the Centre for Digitalisation and Technology Research of the German Armed Forced (DTEC.Bw).
TP acknowledges support of the project CAAS CZ.02.1.01\/0.0\/0.0\/16\_019\/0000778.
\end{acknowledgments}

\section*{Author contributions statement}
A.S., J.C., and DK performed the simulations. A.S. and D.K. conceptualized the study and wrote the manuscripts. D.K. and T.P. supervised the work. All authors reviewed the manuscript. 

\section*{Competing Interests}
The authors declare no competing interests

% \begin{suppinfo}
% \end{suppinfo}

%\bibliography{biblio.bib}

\begin{thebibliography}{61}%
\makeatletter
\providecommand \@ifxundefined [1]{%
 \@ifx{#1\undefined}
}%
\providecommand \@ifnum [1]{%
 \ifnum #1\expandafter \@firstoftwo
 \else \expandafter \@secondoftwo
 \fi
}%
\providecommand \@ifx [1]{%
 \ifx #1\expandafter \@firstoftwo
 \else \expandafter \@secondoftwo
 \fi
}%
\providecommand \natexlab [1]{#1}%
\providecommand \enquote  [1]{``#1''}%
\providecommand \bibnamefont  [1]{#1}%
\providecommand \bibfnamefont [1]{#1}%
\providecommand \citenamefont [1]{#1}%
\providecommand \href@noop [0]{\@secondoftwo}%
\providecommand \href [0]{\begingroup \@sanitize@url \@href}%
\providecommand \@href[1]{\@@startlink{#1}\@@href}%
\providecommand \@@href[1]{\endgroup#1\@@endlink}%
\providecommand \@sanitize@url [0]{\catcode `\\12\catcode `\$12\catcode
  `\&12\catcode `\#12\catcode `\^12\catcode `\_12\catcode `\%12\relax}%
\providecommand \@@startlink[1]{}%
\providecommand \@@endlink[0]{}%
\providecommand \url  [0]{\begingroup\@sanitize@url \@url }%
\providecommand \@url [1]{\endgroup\@href {#1}{\urlprefix }}%
\providecommand \urlprefix  [0]{URL }%
\providecommand \Eprint [0]{\href }%
\providecommand \doibase [0]{https://doi.org/}%
\providecommand \selectlanguage [0]{\@gobble}%
\providecommand \bibinfo  [0]{\@secondoftwo}%
\providecommand \bibfield  [0]{\@secondoftwo}%
\providecommand \translation [1]{[#1]}%
\providecommand \BibitemOpen [0]{}%
\providecommand \bibitemStop [0]{}%
\providecommand \bibitemNoStop [0]{.\EOS\space}%
\providecommand \EOS [0]{\spacefactor3000\relax}%
\providecommand \BibitemShut  [1]{\csname bibitem#1\endcsname}%
\let\auto@bib@innerbib\@empty
%</preamble>
\bibitem [{\citenamefont {Smole{\'{n}}ski}\ \emph {et~al.}(2020)\citenamefont
  {Smole{\'{n}}ski}, \citenamefont {Dolgirev}, \citenamefont {Kuhlenkamp},
  \citenamefont {Popert}, \citenamefont {Shimazaki}, \citenamefont {Back},
  \citenamefont {Kroner}, \citenamefont {Watanabe}, \citenamefont {Taniguchi},
  \citenamefont {Esterlis}, \citenamefont {Demler},\ and\ \citenamefont
  {Imamo{\u{g}}lu}}]{Smolenski2020}%
  \BibitemOpen
  \bibfield  {author} {\bibinfo {author} {\bibfnamefont {T.}~\bibnamefont
  {Smole{\'{n}}ski}}, \bibinfo {author} {\bibfnamefont {P.~E.}\ \bibnamefont
  {Dolgirev}}, \bibinfo {author} {\bibfnamefont {C.}~\bibnamefont
  {Kuhlenkamp}}, \bibinfo {author} {\bibfnamefont {A.}~\bibnamefont {Popert}},
  \bibinfo {author} {\bibfnamefont {Y.}~\bibnamefont {Shimazaki}}, \bibinfo
  {author} {\bibfnamefont {P.}~\bibnamefont {Back}}, \bibinfo {author}
  {\bibfnamefont {M.}~\bibnamefont {Kroner}}, \bibinfo {author} {\bibfnamefont
  {K.}~\bibnamefont {Watanabe}}, \bibinfo {author} {\bibfnamefont
  {T.}~\bibnamefont {Taniguchi}}, \bibinfo {author} {\bibfnamefont
  {I.}~\bibnamefont {Esterlis}}, \bibinfo {author} {\bibfnamefont
  {E.}~\bibnamefont {Demler}},\ and\ \bibinfo {author} {\bibfnamefont
  {A.}~\bibnamefont {Imamo{\u{g}}lu}},\ }\bibfield  {title} {\bibinfo {title}
  {{Observation of Wigner crystal of electrons in a monolayer semiconductor}},\
  }\href {http://arxiv.org/abs/2010.03078} {\bibfield  {journal} {\bibinfo
  {journal} {Nature}\ }\textbf {\bibinfo {volume} {595}} (\bibinfo {year}
  {2020})}\BibitemShut {NoStop}%
\bibitem [{\citenamefont {Song}\ \emph {et~al.}(2018)\citenamefont {Song},
  \citenamefont {Mandelli}, \citenamefont {Hod}, \citenamefont {Urbakh},
  \citenamefont {Ma},\ and\ \citenamefont {Zheng}}]{Song2018RobustSuperlub}%
  \BibitemOpen
  \bibfield  {author} {\bibinfo {author} {\bibfnamefont {Y.}~\bibnamefont
  {Song}}, \bibinfo {author} {\bibfnamefont {D.}~\bibnamefont {Mandelli}},
  \bibinfo {author} {\bibfnamefont {O.}~\bibnamefont {Hod}}, \bibinfo {author}
  {\bibfnamefont {M.}~\bibnamefont {Urbakh}}, \bibinfo {author} {\bibfnamefont
  {M.}~\bibnamefont {Ma}},\ and\ \bibinfo {author} {\bibfnamefont
  {Q.}~\bibnamefont {Zheng}},\ }\bibfield  {title} {\bibinfo {title} {{Robust
  microscale superlubricity in graphite/hexagonal boron nitride layered
  heterojunctions}},\ }\href {https://doi.org/10.1038/s41563-018-0144-z}
  {\bibfield  {journal} {\bibinfo  {journal} {Nature Materials}\ }\textbf
  {\bibinfo {volume} {17}},\ \bibinfo {pages} {894} (\bibinfo {year}
  {2018})}\BibitemShut {NoStop}%
\bibitem [{\citenamefont {Das}\ \emph {et~al.}(2015)\citenamefont {Das},
  \citenamefont {Demarteau},\ and\ \citenamefont {Roelofs}}]{Das2015}%
  \BibitemOpen
  \bibfield  {author} {\bibinfo {author} {\bibfnamefont {S.}~\bibnamefont
  {Das}}, \bibinfo {author} {\bibfnamefont {M.}~\bibnamefont {Demarteau}},\
  and\ \bibinfo {author} {\bibfnamefont {A.}~\bibnamefont {Roelofs}},\
  }\bibfield  {title} {\bibinfo {title} {{Nb-doped single crystalline MoS$_{2}$
  field effect transistor}},\ }\bibfield  {journal} {\bibinfo  {journal}
  {Applied Physics Letters}\ }\textbf {\bibinfo {volume} {106}},\ \href
  {https://doi.org/10.1063/1.4919565} {10.1063/1.4919565} (\bibinfo {year}
  {2015})\BibitemShut {NoStop}%
\bibitem [{\citenamefont {Pattengale}\ \emph {et~al.}(2020)\citenamefont
  {Pattengale}, \citenamefont {Huang}, \citenamefont {Yan}, \citenamefont
  {Yang}, \citenamefont {Younan}, \citenamefont {Hu}, \citenamefont {Li},
  \citenamefont {Lee}, \citenamefont {Pan}, \citenamefont {Gu},\ and\
  \citenamefont {Huang}}]{Pattengale2020}%
  \BibitemOpen
  \bibfield  {author} {\bibinfo {author} {\bibfnamefont {B.}~\bibnamefont
  {Pattengale}}, \bibinfo {author} {\bibfnamefont {Y.}~\bibnamefont {Huang}},
  \bibinfo {author} {\bibfnamefont {X.}~\bibnamefont {Yan}}, \bibinfo {author}
  {\bibfnamefont {S.}~\bibnamefont {Yang}}, \bibinfo {author} {\bibfnamefont
  {S.}~\bibnamefont {Younan}}, \bibinfo {author} {\bibfnamefont
  {W.}~\bibnamefont {Hu}}, \bibinfo {author} {\bibfnamefont {Z.}~\bibnamefont
  {Li}}, \bibinfo {author} {\bibfnamefont {S.}~\bibnamefont {Lee}}, \bibinfo
  {author} {\bibfnamefont {X.}~\bibnamefont {Pan}}, \bibinfo {author}
  {\bibfnamefont {J.}~\bibnamefont {Gu}},\ and\ \bibinfo {author}
  {\bibfnamefont {J.}~\bibnamefont {Huang}},\ }\bibfield  {title} {\bibinfo
  {title} {{Dynamic evolution and reversibility of single-atom Ni(II) active
  site in 1T-MoS$_2$ electrocatalysts for hydrogen evolution}},\ }\href
  {https://doi.org/10.1038/s41467-020-17904-z} {\bibfield  {journal} {\bibinfo
  {journal} {Nature Communications}\ }\textbf {\bibinfo {volume} {11}},\
  \bibinfo {pages} {4114} (\bibinfo {year} {2020})}\BibitemShut {NoStop}%
\bibitem [{\citenamefont {Mounet}\ \emph {et~al.}(2018)\citenamefont {Mounet},
  \citenamefont {Gibertini}, \citenamefont {Schwaller}, \citenamefont {Campi},
  \citenamefont {Merkys}, \citenamefont {Marrazzo}, \citenamefont {Sohier},
  \citenamefont {Castelli}, \citenamefont {Cepellotti}, \citenamefont {Pizzi},\
  and\ \citenamefont {Marzari}}]{Mounet2018}%
  \BibitemOpen
  \bibfield  {author} {\bibinfo {author} {\bibfnamefont {N.}~\bibnamefont
  {Mounet}}, \bibinfo {author} {\bibfnamefont {M.}~\bibnamefont {Gibertini}},
  \bibinfo {author} {\bibfnamefont {P.}~\bibnamefont {Schwaller}}, \bibinfo
  {author} {\bibfnamefont {D.}~\bibnamefont {Campi}}, \bibinfo {author}
  {\bibfnamefont {A.}~\bibnamefont {Merkys}}, \bibinfo {author} {\bibfnamefont
  {A.}~\bibnamefont {Marrazzo}}, \bibinfo {author} {\bibfnamefont
  {T.}~\bibnamefont {Sohier}}, \bibinfo {author} {\bibfnamefont {I.~E.}\
  \bibnamefont {Castelli}}, \bibinfo {author} {\bibfnamefont {A.}~\bibnamefont
  {Cepellotti}}, \bibinfo {author} {\bibfnamefont {G.}~\bibnamefont {Pizzi}},\
  and\ \bibinfo {author} {\bibfnamefont {N.}~\bibnamefont {Marzari}},\
  }\bibfield  {title} {\bibinfo {title} {{Two-dimensional materials from
  high-throughput computational exfoliation of experimentally known
  compounds}},\ }\href {https://doi.org/10.1038/s41565-017-0035-5} {\bibfield
  {journal} {\bibinfo  {journal} {Nature Nanotechnology}\ }\textbf {\bibinfo
  {volume} {13}},\ \bibinfo {pages} {246} (\bibinfo {year} {2018})}\BibitemShut
  {NoStop}%
\bibitem [{\citenamefont {Sorkun}\ \emph {et~al.}(2020)\citenamefont {Sorkun},
  \citenamefont {Astruc}, \citenamefont {Koelman},\ and\ \citenamefont
  {Er}}]{Sorkun2020}%
  \BibitemOpen
  \bibfield  {author} {\bibinfo {author} {\bibfnamefont {M.~C.}\ \bibnamefont
  {Sorkun}}, \bibinfo {author} {\bibfnamefont {S.}~\bibnamefont {Astruc}},
  \bibinfo {author} {\bibfnamefont {J.~M.~A.}\ \bibnamefont {Koelman}},\ and\
  \bibinfo {author} {\bibfnamefont {S.}~\bibnamefont {Er}},\ }\bibfield
  {title} {\bibinfo {title} {{An artificial intelligence-aided virtual
  screening recipe for two-dimensional materials discovery}},\ }\href
  {https://doi.org/10.1038/s41524-020-00375-7} {\bibfield  {journal} {\bibinfo
  {journal} {npj Computational Materials}\ }\textbf {\bibinfo {volume} {6}},\
  \bibinfo {pages} {1} (\bibinfo {year} {2020})}\BibitemShut {NoStop}%
\bibitem [{\citenamefont {Kumar}\ \emph {et~al.}(2022)\citenamefont {Kumar},
  \citenamefont {Sharma}, \citenamefont {Shirodkar},\ and\ \citenamefont
  {Dev}}]{Kumar2022TMD_ML}%
  \BibitemOpen
  \bibfield  {author} {\bibinfo {author} {\bibfnamefont {P.}~\bibnamefont
  {Kumar}}, \bibinfo {author} {\bibfnamefont {V.}~\bibnamefont {Sharma}},
  \bibinfo {author} {\bibfnamefont {S.~N.}\ \bibnamefont {Shirodkar}},\ and\
  \bibinfo {author} {\bibfnamefont {P.}~\bibnamefont {Dev}},\ }\bibfield
  {title} {\bibinfo {title} {{Predicting phase preferences of two-dimensional
  transition metal dichalcogenides using machine learning}},\ }\href
  {https://doi.org/10.1103/PhysRevMaterials.6.094007} {\bibfield  {journal}
  {\bibinfo  {journal} {Physical Review Materials}\ }\textbf {\bibinfo {volume}
  {6}},\ \bibinfo {pages} {094007} (\bibinfo {year} {2022})}\BibitemShut
  {NoStop}%
\bibitem [{\citenamefont {Zhou}\ \emph {et~al.}(2018)\citenamefont {Zhou},
  \citenamefont {Lin}, \citenamefont {Huang}, \citenamefont {Zhou},
  \citenamefont {Chen}, \citenamefont {Xia}, \citenamefont {Wang},
  \citenamefont {Xie}, \citenamefont {Yu}, \citenamefont {Lei}, \citenamefont
  {Wu}, \citenamefont {Liu}, \citenamefont {Fu}, \citenamefont {Zeng},
  \citenamefont {Hsu}, \citenamefont {Yang}, \citenamefont {Lu}, \citenamefont
  {Yu}, \citenamefont {Shen}, \citenamefont {Lin}, \citenamefont {Yakobson},
  \citenamefont {Liu}, \citenamefont {Suenaga}, \citenamefont {Liu},\ and\
  \citenamefont {Liu}}]{Zhou2018}%
  \BibitemOpen
  \bibfield  {author} {\bibinfo {author} {\bibfnamefont {J.}~\bibnamefont
  {Zhou}}, \bibinfo {author} {\bibfnamefont {J.}~\bibnamefont {Lin}}, \bibinfo
  {author} {\bibfnamefont {X.}~\bibnamefont {Huang}}, \bibinfo {author}
  {\bibfnamefont {Y.}~\bibnamefont {Zhou}}, \bibinfo {author} {\bibfnamefont
  {Y.}~\bibnamefont {Chen}}, \bibinfo {author} {\bibfnamefont {J.}~\bibnamefont
  {Xia}}, \bibinfo {author} {\bibfnamefont {H.}~\bibnamefont {Wang}}, \bibinfo
  {author} {\bibfnamefont {Y.}~\bibnamefont {Xie}}, \bibinfo {author}
  {\bibfnamefont {H.}~\bibnamefont {Yu}}, \bibinfo {author} {\bibfnamefont
  {J.}~\bibnamefont {Lei}}, \bibinfo {author} {\bibfnamefont {D.}~\bibnamefont
  {Wu}}, \bibinfo {author} {\bibfnamefont {F.}~\bibnamefont {Liu}}, \bibinfo
  {author} {\bibfnamefont {Q.}~\bibnamefont {Fu}}, \bibinfo {author}
  {\bibfnamefont {Q.}~\bibnamefont {Zeng}}, \bibinfo {author} {\bibfnamefont
  {C.~H.}\ \bibnamefont {Hsu}}, \bibinfo {author} {\bibfnamefont
  {C.}~\bibnamefont {Yang}}, \bibinfo {author} {\bibfnamefont {L.}~\bibnamefont
  {Lu}}, \bibinfo {author} {\bibfnamefont {T.}~\bibnamefont {Yu}}, \bibinfo
  {author} {\bibfnamefont {Z.}~\bibnamefont {Shen}}, \bibinfo {author}
  {\bibfnamefont {H.}~\bibnamefont {Lin}}, \bibinfo {author} {\bibfnamefont
  {B.~I.}\ \bibnamefont {Yakobson}}, \bibinfo {author} {\bibfnamefont
  {Q.}~\bibnamefont {Liu}}, \bibinfo {author} {\bibfnamefont {K.}~\bibnamefont
  {Suenaga}}, \bibinfo {author} {\bibfnamefont {G.}~\bibnamefont {Liu}},\ and\
  \bibinfo {author} {\bibfnamefont {Z.}~\bibnamefont {Liu}},\ }\bibfield
  {title} {\bibinfo {title} {{A library of atomically thin metal
  chalcogenides}},\ }\href {https://doi.org/10.1038/s41586-018-0008-3}
  {\bibfield  {journal} {\bibinfo  {journal} {Nature}\ }\textbf {\bibinfo
  {volume} {556}},\ \bibinfo {pages} {355} (\bibinfo {year}
  {2018})}\BibitemShut {NoStop}%
\bibitem [{\citenamefont {Shivayogimath}\ \emph {et~al.}(2019)\citenamefont
  {Shivayogimath}, \citenamefont {Thomsen}, \citenamefont {Mackenzie},
  \citenamefont {Geisler}, \citenamefont {Stan}, \citenamefont {Holt},
  \citenamefont {Bianchi}, \citenamefont {Crovetto}, \citenamefont {Whelan},
  \citenamefont {Carvalho}, \citenamefont {Neto}, \citenamefont {Hofmann},
  \citenamefont {Stenger}, \citenamefont {B{\o}ggild},\ and\ \citenamefont
  {Booth}}]{Shivayogimath2018}%
  \BibitemOpen
  \bibfield  {author} {\bibinfo {author} {\bibfnamefont {A.}~\bibnamefont
  {Shivayogimath}}, \bibinfo {author} {\bibfnamefont {J.~D.}\ \bibnamefont
  {Thomsen}}, \bibinfo {author} {\bibfnamefont {D.~M.~A.}\ \bibnamefont
  {Mackenzie}}, \bibinfo {author} {\bibfnamefont {M.}~\bibnamefont {Geisler}},
  \bibinfo {author} {\bibfnamefont {R.-M.}\ \bibnamefont {Stan}}, \bibinfo
  {author} {\bibfnamefont {A.~J.}\ \bibnamefont {Holt}}, \bibinfo {author}
  {\bibfnamefont {M.}~\bibnamefont {Bianchi}}, \bibinfo {author} {\bibfnamefont
  {A.}~\bibnamefont {Crovetto}}, \bibinfo {author} {\bibfnamefont {P.~R.}\
  \bibnamefont {Whelan}}, \bibinfo {author} {\bibfnamefont {A.}~\bibnamefont
  {Carvalho}}, \bibinfo {author} {\bibfnamefont {A.~H.~C.}\ \bibnamefont
  {Neto}}, \bibinfo {author} {\bibfnamefont {P.}~\bibnamefont {Hofmann}},
  \bibinfo {author} {\bibfnamefont {N.}~\bibnamefont {Stenger}}, \bibinfo
  {author} {\bibfnamefont {P.}~\bibnamefont {B{\o}ggild}},\ and\ \bibinfo
  {author} {\bibfnamefont {T.~J.}\ \bibnamefont {Booth}},\ }\bibfield  {title}
  {\bibinfo {title} {{A universal approach for the synthesis of two-dimensional
  binary compounds}},\ }\href {https://doi.org/10.1038/s41467-019-11075-2}
  {\bibfield  {journal} {\bibinfo  {journal} {Nature Communications}\ }\textbf
  {\bibinfo {volume} {10}},\ \bibinfo {pages} {2957} (\bibinfo {year}
  {2019})}\BibitemShut {NoStop}%
\bibitem [{\citenamefont {Domask}\ \emph {et~al.}(2015)\citenamefont {Domask},
  \citenamefont {Gurunathan},\ and\ \citenamefont {Mohney}}]{Domask2015}%
  \BibitemOpen
  \bibfield  {author} {\bibinfo {author} {\bibfnamefont {A.~C.}\ \bibnamefont
  {Domask}}, \bibinfo {author} {\bibfnamefont {R.~L.}\ \bibnamefont
  {Gurunathan}},\ and\ \bibinfo {author} {\bibfnamefont {S.~E.}\ \bibnamefont
  {Mohney}},\ }\bibfield  {title} {\bibinfo {title} {Transition metal-mos$_2$
  reactions: Review and thermodynamic predictions},\ }\href
  {https://doi.org/10.1007/s11664-015-3956-5} {\bibfield  {journal} {\bibinfo
  {journal} {Journal of Electronic Materials}\ }\textbf {\bibinfo {volume}
  {44}},\ \bibinfo {pages} {4065} (\bibinfo {year} {2015})}\BibitemShut
  {NoStop}%
\bibitem [{\citenamefont {Woods-Robinson}\ \emph {et~al.}()\citenamefont
  {Woods-Robinson}, \citenamefont {Horton},\ and\ \citenamefont
  {Persson}}]{Woods-Robinson2022alloy_screening}%
  \BibitemOpen
  \bibfield  {author} {\bibinfo {author} {\bibfnamefont {R.}~\bibnamefont
  {Woods-Robinson}}, \bibinfo {author} {\bibfnamefont {M.~K.}\ \bibnamefont
  {Horton}},\ and\ \bibinfo {author} {\bibfnamefont {K.~A.}\ \bibnamefont
  {Persson}},\ }\bibfield  {title} {\bibinfo {title} {{A method to
  computationally screen for tunable properties of crystalline alloys}},\
  }\href@noop {} {\bibinfo  {journal} {(21 Jun 2022) ArXiv (Condensed Matter,
  Materials Science) DOI: 10.48550/arXiv.2206.10715 (accessed: 2022-11-05)}\
  }\BibitemShut {NoStop}%
\bibitem [{\citenamefont {Koepernik}\ \emph {et~al.}(2016)\citenamefont
  {Koepernik}, \citenamefont {Kasinathan}, \citenamefont {Efremov},
  \citenamefont {Khim}, \citenamefont {Borisenko}, \citenamefont
  {B{\"{u}}chner},\ and\ \citenamefont {Van Den~Brink}}]{Koepernik2016}%
  \BibitemOpen
\bibfield  {journal} {  }\bibfield  {author} {\bibinfo {author} {\bibfnamefont
  {K.}~\bibnamefont {Koepernik}}, \bibinfo {author} {\bibfnamefont
  {D.}~\bibnamefont {Kasinathan}}, \bibinfo {author} {\bibfnamefont {D.~V.}\
  \bibnamefont {Efremov}}, \bibinfo {author} {\bibfnamefont {S.}~\bibnamefont
  {Khim}}, \bibinfo {author} {\bibfnamefont {S.}~\bibnamefont {Borisenko}},
  \bibinfo {author} {\bibfnamefont {B.}~\bibnamefont {B{\"{u}}chner}},\ and\
  \bibinfo {author} {\bibfnamefont {J.}~\bibnamefont {Van Den~Brink}},\
  }\bibfield  {title} {\bibinfo {title} {{TaIrTe$_{4}$: A ternary type-II Weyl
  semimetal}},\ }\href {https://doi.org/10.1103/PhysRevB.93.201101} {\bibfield
  {journal} {\bibinfo  {journal} {Physical Review B}\ }\textbf {\bibinfo
  {volume} {93}},\ \bibinfo {pages} {1} (\bibinfo {year} {2016})}\BibitemShut
  {NoStop}%
\bibitem [{\citenamefont {Saeki}\ and\ \citenamefont
  {Onoda}(1987)}]{SAEKI1987}%
  \BibitemOpen
  \bibfield  {author} {\bibinfo {author} {\bibfnamefont {M.}~\bibnamefont
  {Saeki}}\ and\ \bibinfo {author} {\bibfnamefont {M.}~\bibnamefont {Onoda}},\
  }\bibfield  {title} {\bibinfo {title} {Preparation of 3s-type
  {Mo}$_{0.5}${Ta}$_{0.5}${S}$_{2}$},\ }\href
  {https://pascal-francis.inist.fr/vibad/index.php?action=getRecordDetail&idt=7652667}
  {\bibfield  {journal} {\bibinfo  {journal} {Journal of the less-common
  metals}\ }\textbf {\bibinfo {volume} {135}},\ \bibinfo {pages} {L1} (\bibinfo
  {year} {1987})}\BibitemShut {NoStop}%
\bibitem [{\citenamefont {Gao}\ \emph {et~al.}(2020)\citenamefont {Gao},
  \citenamefont {Gao}, \citenamefont {Suh}, \citenamefont {Suh}, \citenamefont
  {Cao}, \citenamefont {Joe}, \citenamefont {Mujid}, \citenamefont {Lee},
  \citenamefont {Lee}, \citenamefont {Xie}, \citenamefont {Xie}, \citenamefont
  {Poddar}, \citenamefont {Lee}, \citenamefont {Lee}, \citenamefont {Kang},
  \citenamefont {Kang}, \citenamefont {Kim}, \citenamefont {Muller},\ and\
  \citenamefont {Park}}]{Gao2020}%
  \BibitemOpen
  \bibfield  {author} {\bibinfo {author} {\bibfnamefont {H.}~\bibnamefont
  {Gao}}, \bibinfo {author} {\bibfnamefont {H.}~\bibnamefont {Gao}}, \bibinfo
  {author} {\bibfnamefont {J.}~\bibnamefont {Suh}}, \bibinfo {author}
  {\bibfnamefont {J.}~\bibnamefont {Suh}}, \bibinfo {author} {\bibfnamefont
  {M.~C.}\ \bibnamefont {Cao}}, \bibinfo {author} {\bibfnamefont {A.~Y.}\
  \bibnamefont {Joe}}, \bibinfo {author} {\bibfnamefont {F.}~\bibnamefont
  {Mujid}}, \bibinfo {author} {\bibfnamefont {K.~H.}\ \bibnamefont {Lee}},
  \bibinfo {author} {\bibfnamefont {K.~H.}\ \bibnamefont {Lee}}, \bibinfo
  {author} {\bibfnamefont {S.}~\bibnamefont {Xie}}, \bibinfo {author}
  {\bibfnamefont {S.}~\bibnamefont {Xie}}, \bibinfo {author} {\bibfnamefont
  {P.}~\bibnamefont {Poddar}}, \bibinfo {author} {\bibfnamefont {J.~U.}\
  \bibnamefont {Lee}}, \bibinfo {author} {\bibfnamefont {J.~U.}\ \bibnamefont
  {Lee}}, \bibinfo {author} {\bibfnamefont {K.}~\bibnamefont {Kang}}, \bibinfo
  {author} {\bibfnamefont {K.}~\bibnamefont {Kang}}, \bibinfo {author}
  {\bibfnamefont {P.}~\bibnamefont {Kim}}, \bibinfo {author} {\bibfnamefont
  {D.~A.}\ \bibnamefont {Muller}},\ and\ \bibinfo {author} {\bibfnamefont
  {J.}~\bibnamefont {Park}},\ }\bibfield  {title} {\bibinfo {title} {Tuning
  electrical conductance of {MoS}$_{2}$ monolayers through substitutional
  doping},\ }\href {https://doi.org/10.1021/acs.nanolett.9b05247} {\bibfield
  {journal} {\bibinfo  {journal} {Nano Letters}\ }\textbf {\bibinfo {volume}
  {20}},\ \bibinfo {pages} {4095} (\bibinfo {year} {2020})}\BibitemShut
  {NoStop}%
\bibitem [{\citenamefont {Han}\ \emph {et~al.}(2020)\citenamefont {Han},
  \citenamefont {Benkraouda}, \citenamefont {Qamhieh},\ and\ \citenamefont
  {Amrane}}]{Han2020a}%
  \BibitemOpen
  \bibfield  {author} {\bibinfo {author} {\bibfnamefont {X.}~\bibnamefont
  {Han}}, \bibinfo {author} {\bibfnamefont {M.}~\bibnamefont {Benkraouda}},
  \bibinfo {author} {\bibfnamefont {N.}~\bibnamefont {Qamhieh}},\ and\ \bibinfo
  {author} {\bibfnamefont {N.}~\bibnamefont {Amrane}},\ }\bibfield  {title}
  {\bibinfo {title} {{Understanding ferromagnetism in Ni-doped MoS$_{2}$
  monolayer from first principles}},\ }\href
  {https://doi.org/10.1016/j.chemphys.2019.110501} {\bibfield  {journal}
  {\bibinfo  {journal} {Chemical Physics}\ }\textbf {\bibinfo {volume} {528}},\
  \bibinfo {pages} {110501} (\bibinfo {year} {2020})}\BibitemShut {NoStop}%
\bibitem [{\citenamefont {Chen}\ \emph {et~al.}(2013)\citenamefont {Chen},
  \citenamefont {Xi}, \citenamefont {Dumcenco}, \citenamefont {Liu},
  \citenamefont {Suenaga}, \citenamefont {Wang}, \citenamefont {Shuai},
  \citenamefont {Huang},\ and\ \citenamefont {Xie}}]{Chen2013}%
  \BibitemOpen
  \bibfield  {author} {\bibinfo {author} {\bibfnamefont {Y.}~\bibnamefont
  {Chen}}, \bibinfo {author} {\bibfnamefont {J.}~\bibnamefont {Xi}}, \bibinfo
  {author} {\bibfnamefont {D.~O.}\ \bibnamefont {Dumcenco}}, \bibinfo {author}
  {\bibfnamefont {Z.}~\bibnamefont {Liu}}, \bibinfo {author} {\bibfnamefont
  {K.}~\bibnamefont {Suenaga}}, \bibinfo {author} {\bibfnamefont
  {D.}~\bibnamefont {Wang}}, \bibinfo {author} {\bibfnamefont {Z.}~\bibnamefont
  {Shuai}}, \bibinfo {author} {\bibfnamefont {Y.~S.}\ \bibnamefont {Huang}},\
  and\ \bibinfo {author} {\bibfnamefont {L.}~\bibnamefont {Xie}},\ }\bibfield
  {title} {\bibinfo {title} {{Tunable band gap photoluminescence from
  atomically thin transition-metal dichalcogenide alloys}},\ }\href
  {https://doi.org/10.1021/nn401420h} {\bibfield  {journal} {\bibinfo
  {journal} {ACS Nano}\ }\textbf {\bibinfo {volume} {7}},\ \bibinfo {pages}
  {4610} (\bibinfo {year} {2013})}\BibitemShut {NoStop}%
\bibitem [{\citenamefont {Worsdale}\ \emph {et~al.}(2015)\citenamefont
  {Worsdale}, \citenamefont {Rabis}, \citenamefont {Fabbri}, \citenamefont
  {Schmidt},\ and\ \citenamefont {Kramer}}]{Worsdale2015a}%
  \BibitemOpen
  \bibfield  {author} {\bibinfo {author} {\bibfnamefont {M.}~\bibnamefont
  {Worsdale}}, \bibinfo {author} {\bibfnamefont {A.}~\bibnamefont {Rabis}},
  \bibinfo {author} {\bibfnamefont {E.}~\bibnamefont {Fabbri}}, \bibinfo
  {author} {\bibfnamefont {T.~J.}\ \bibnamefont {Schmidt}},\ and\ \bibinfo
  {author} {\bibfnamefont {D.}~\bibnamefont {Kramer}},\ }\bibfield  {title}
  {\bibinfo {title} {{Conductivity Limits of Extrinsically Doped SnO$_{2}$
  Supports}},\ }\href {https://doi.org/10.1149/06917.1167ecst} {\bibfield
  {journal} {\bibinfo  {journal} {ECS Transactions}\ }\textbf {\bibinfo
  {volume} {69}},\ \bibinfo {pages} {1167} (\bibinfo {year}
  {2015})}\BibitemShut {NoStop}%
\bibitem [{\citenamefont {Abbott}(1934)}]{Abbott}%
  \BibitemOpen
  \bibfield  {author} {\bibinfo {author} {\bibfnamefont {G.~W.~M.}\
  \bibnamefont {Abbott}},\ }\bibfield  {title} {\bibinfo {title} {{The freezing
  points, melting points, and solid solubility limits of the alloys of sliver
  and copper with the elements of the b sub-groups}},\ }\href
  {https://doi.org/10.1098/rsta.1934.0014} {\bibfield  {journal} {\bibinfo
  {journal} {Philosophical Transactions of the Royal Society of London. Series
  A, Containing Papers of a Mathematical or Physical Character}\ }\textbf
  {\bibinfo {volume} {233}},\ \bibinfo {pages} {1} (\bibinfo {year}
  {1934})}\BibitemShut {NoStop}%
\bibitem [{\citenamefont {Pettifor}(1986)}]{Pettifor1986}%
  \BibitemOpen
  \bibfield  {author} {\bibinfo {author} {\bibfnamefont {D.~G.}\ \bibnamefont
  {Pettifor}},\ }\bibfield  {title} {\bibinfo {title} {{The structures of
  binary compounds: I. phenomenological structure maps}},\ }\href
  {https://doi.org/10.1088/0022-3719/19/3/002} {\bibfield  {journal} {\bibinfo
  {journal} {Journal of Physics C: Solid State Physics}\ }\textbf {\bibinfo
  {volume} {19}},\ \bibinfo {pages} {285} (\bibinfo {year} {1986})}\BibitemShut
  {NoStop}%
\bibitem [{\citenamefont {Connolly}\ and\ \citenamefont
  {Williams}(1983)}]{Connolly1983}%
  \BibitemOpen
  \bibfield  {author} {\bibinfo {author} {\bibfnamefont {J.~W.~D.}\
  \bibnamefont {Connolly}}\ and\ \bibinfo {author} {\bibfnamefont {A.~R.}\
  \bibnamefont {Williams}},\ }\bibfield  {title} {\bibinfo {title}
  {{Density-functional theory applied to phase transformations in
  transition-metal alloys}},\ }\href {https://doi.org/10.1103/PhysRevB.27.5169}
  {\bibfield  {journal} {\bibinfo  {journal} {Physical Review B}\ }\textbf
  {\bibinfo {volume} {27}},\ \bibinfo {pages} {5169} (\bibinfo {year}
  {1983})}\BibitemShut {NoStop}%
\bibitem [{\citenamefont {Hautier}\ \emph {et~al.}(2011)\citenamefont
  {Hautier}, \citenamefont {Fischer}, \citenamefont {Ehrlacher}, \citenamefont
  {Jain},\ and\ \citenamefont {Ceder}}]{Hautier2011}%
  \BibitemOpen
  \bibfield  {author} {\bibinfo {author} {\bibfnamefont {G.}~\bibnamefont
  {Hautier}}, \bibinfo {author} {\bibfnamefont {C.}~\bibnamefont {Fischer}},
  \bibinfo {author} {\bibfnamefont {V.}~\bibnamefont {Ehrlacher}}, \bibinfo
  {author} {\bibfnamefont {A.}~\bibnamefont {Jain}},\ and\ \bibinfo {author}
  {\bibfnamefont {G.}~\bibnamefont {Ceder}},\ }\bibfield  {title} {\bibinfo
  {title} {{Data mined ionic substitutions for the discovery of new
  compounds}},\ }\href {https://doi.org/10.1021/ic102031h} {\bibfield
  {journal} {\bibinfo  {journal} {Inorganic Chemistry}\ }\textbf {\bibinfo
  {volume} {50}},\ \bibinfo {pages} {656} (\bibinfo {year} {2011})}\BibitemShut
  {NoStop}%
\bibitem [{\citenamefont {Ceder}\ \emph {et~al.}(2000)\citenamefont {Ceder},
  \citenamefont {Van Der~Ven}, \citenamefont {Marianetti},\ and\ \citenamefont
  {Morgan}}]{Ceder2000}%
  \BibitemOpen
  \bibfield  {author} {\bibinfo {author} {\bibfnamefont {G.}~\bibnamefont
  {Ceder}}, \bibinfo {author} {\bibfnamefont {A.}~\bibnamefont {Van Der~Ven}},
  \bibinfo {author} {\bibfnamefont {C.}~\bibnamefont {Marianetti}},\ and\
  \bibinfo {author} {\bibfnamefont {D.}~\bibnamefont {Morgan}},\ }\bibfield
  {title} {\bibinfo {title} {{First-principles alloy theory in oxides}},\
  }\href {https://doi.org/10.1088/0965-0393/8/3/311} {\bibfield  {journal}
  {\bibinfo  {journal} {Modelling and Simulation in Materials Science and
  Engineering}\ }\textbf {\bibinfo {volume} {8}},\ \bibinfo {pages} {311}
  (\bibinfo {year} {2000})}\BibitemShut {NoStop}%
\bibitem [{\citenamefont {Furlan}\ \emph {et~al.}(2015)\citenamefont {Furlan},
  \citenamefont {Prates}, \citenamefont {Andrea Dos~Santos}, \citenamefont
  {Gouv{\^{e}}a~Dias}, \citenamefont {Ferreira}, \citenamefont
  {Rodrigues~Neto},\ and\ \citenamefont {Klein}}]{Furlan2015}%
  \BibitemOpen
  \bibfield  {author} {\bibinfo {author} {\bibfnamefont {K.~P.}\ \bibnamefont
  {Furlan}}, \bibinfo {author} {\bibfnamefont {P.~B.}\ \bibnamefont {Prates}},
  \bibinfo {author} {\bibfnamefont {T.}~\bibnamefont {Andrea Dos~Santos}},
  \bibinfo {author} {\bibfnamefont {M.~V.}\ \bibnamefont {Gouv{\^{e}}a~Dias}},
  \bibinfo {author} {\bibfnamefont {H.~T.}\ \bibnamefont {Ferreira}}, \bibinfo
  {author} {\bibfnamefont {J.~B.}\ \bibnamefont {Rodrigues~Neto}},\ and\
  \bibinfo {author} {\bibfnamefont {A.~N.}\ \bibnamefont {Klein}},\ }\bibfield
  {title} {\bibinfo {title} {{Influence of alloying elements on the sintering
  thermodynamics, microstructure and properties of Fe-MoS$_{2}$ composites}},\
  }\href {https://doi.org/10.1016/j.jallcom.2015.08.242} {\bibfield  {journal}
  {\bibinfo  {journal} {Journal of Alloys and Compounds}\ }\textbf {\bibinfo
  {volume} {652}},\ \bibinfo {pages} {450} (\bibinfo {year}
  {2015})}\BibitemShut {NoStop}%
\bibitem [{\citenamefont {Onofrio}\ \emph {et~al.}(2017)\citenamefont
  {Onofrio}, \citenamefont {Guzman},\ and\ \citenamefont
  {Strachan}}]{Onofrio2017}%
  \BibitemOpen
  \bibfield  {author} {\bibinfo {author} {\bibfnamefont {N.}~\bibnamefont
  {Onofrio}}, \bibinfo {author} {\bibfnamefont {D.}~\bibnamefont {Guzman}},\
  and\ \bibinfo {author} {\bibfnamefont {A.}~\bibnamefont {Strachan}},\
  }\bibfield  {title} {\bibinfo {title} {{Novel doping alternatives for
  single-layer transition metal dichalcogenides}},\ }\href
  {https://doi.org/10.1063/1.4994997 http://arxiv.org/abs/1703.10745}
  {\bibfield  {journal} {\bibinfo  {journal} {Journal of Applied Physics}\
  }\textbf {\bibinfo {volume} {1221}},\ \bibinfo {pages} {185102} (\bibinfo
  {year} {2017})}\BibitemShut {NoStop}%
\bibitem [{\citenamefont {Avery}\ \emph {et~al.}(2019)\citenamefont {Avery},
  \citenamefont {Wang}, \citenamefont {Proserpio}, \citenamefont {Toher},
  \citenamefont {Oses}, \citenamefont {Gossett}, \citenamefont {Curtarolo},
  \citenamefont {Zurek}, \citenamefont {Davide}, \citenamefont {Toher},
  \citenamefont {Curtarolo}, \citenamefont {Zurek},\ and\ \citenamefont
  {Chimica}}]{Avery2019}%
  \BibitemOpen
  \bibfield  {author} {\bibinfo {author} {\bibfnamefont {P.}~\bibnamefont
  {Avery}}, \bibinfo {author} {\bibfnamefont {X.}~\bibnamefont {Wang}},
  \bibinfo {author} {\bibfnamefont {D.~M.}\ \bibnamefont {Proserpio}}, \bibinfo
  {author} {\bibfnamefont {C.}~\bibnamefont {Toher}}, \bibinfo {author}
  {\bibfnamefont {C.}~\bibnamefont {Oses}}, \bibinfo {author} {\bibfnamefont
  {E.}~\bibnamefont {Gossett}}, \bibinfo {author} {\bibfnamefont
  {S.}~\bibnamefont {Curtarolo}}, \bibinfo {author} {\bibfnamefont
  {E.}~\bibnamefont {Zurek}}, \bibinfo {author} {\bibfnamefont
  {M.}~\bibnamefont {Davide}}, \bibinfo {author} {\bibfnamefont
  {C.}~\bibnamefont {Toher}}, \bibinfo {author} {\bibfnamefont
  {S.}~\bibnamefont {Curtarolo}}, \bibinfo {author} {\bibfnamefont
  {E.}~\bibnamefont {Zurek}},\ and\ \bibinfo {author} {\bibfnamefont
  {D.}~\bibnamefont {Chimica}},\ }\bibfield  {title} {\bibinfo {title}
  {{Predicting Superhard Materials via a Machine Learning Informed Evolutionary
  Structure Search}},\ }\href {http://arxiv.org/abs/1906.05886} {\bibfield
  {journal} {\bibinfo  {journal} {npj Computational Materials}\ } (\bibinfo
  {year} {2019})}\BibitemShut {NoStop}%
\bibitem [{\citenamefont {Wang}\ \emph {et~al.}(2004)\citenamefont {Wang},
  \citenamefont {Curtarolo}, \citenamefont {Jiang}, \citenamefont {Arroyave},
  \citenamefont {Wang}, \citenamefont {Ceder}, \citenamefont {Chen},\ and\
  \citenamefont {Liu}}]{Wang2004Calphad}%
  \BibitemOpen
  \bibfield  {author} {\bibinfo {author} {\bibfnamefont {Y.}~\bibnamefont
  {Wang}}, \bibinfo {author} {\bibfnamefont {S.}~\bibnamefont {Curtarolo}},
  \bibinfo {author} {\bibfnamefont {C.}~\bibnamefont {Jiang}}, \bibinfo
  {author} {\bibfnamefont {R.}~\bibnamefont {Arroyave}}, \bibinfo {author}
  {\bibfnamefont {T.}~\bibnamefont {Wang}}, \bibinfo {author} {\bibfnamefont
  {G.}~\bibnamefont {Ceder}}, \bibinfo {author} {\bibfnamefont {L.-Q.}\
  \bibnamefont {Chen}},\ and\ \bibinfo {author} {\bibfnamefont {Z.-K.}\
  \bibnamefont {Liu}},\ }\bibfield  {title} {\bibinfo {title} {{Ab initio
  lattice stability in comparison with CALPHAD lattice stability}},\ }\href
  {https://doi.org/https://doi.org/10.1016/j.calphad.2004.05.002} {\bibfield
  {journal} {\bibinfo  {journal} {Calphad}\ }\textbf {\bibinfo {volume} {28}},\
  \bibinfo {pages} {79} (\bibinfo {year} {2004})}\BibitemShut {NoStop}%
\bibitem [{\citenamefont {Silva}\ \emph {et~al.}(2022)\citenamefont {Silva},
  \citenamefont {Cao}, \citenamefont {Polcar},\ and\ \citenamefont
  {Kramer}}]{Silva2022pettifor}%
  \BibitemOpen
  \bibfield  {author} {\bibinfo {author} {\bibfnamefont {A.}~\bibnamefont
  {Silva}}, \bibinfo {author} {\bibfnamefont {J.}~\bibnamefont {Cao}}, \bibinfo
  {author} {\bibfnamefont {T.}~\bibnamefont {Polcar}},\ and\ \bibinfo {author}
  {\bibfnamefont {D.}~\bibnamefont {Kramer}},\ }\bibfield  {title} {\bibinfo
  {title} {{Pettifor maps of complex ternary two-dimensional transition metal
  sulfides}},\ }\href {https://doi.org/10.1038/s41524-022-00868-7} {\bibfield
  {journal} {\bibinfo  {journal} {npj Computational Materials}\ }\textbf
  {\bibinfo {volume} {8}} (\bibinfo {year} {2022})}\BibitemShut {NoStop}%
\bibitem [{\citenamefont {Wang}(2021)}]{Wang2021AtomicOutlook}%
  \BibitemOpen
  \bibfield  {author} {\bibinfo {author} {\bibfnamefont {X.}~\bibnamefont
  {Wang}},\ }\bibfield  {title} {\bibinfo {title} {{Atomic Layer Deposition of
  Iron, Cobalt, and Nickel Chalcogenides: Progress and Outlook}},\ }\href
  {https://doi.org/10.1021/acs.chemmater.1c01507} {\bibfield  {journal}
  {\bibinfo  {journal} {Chemistry of Materials}\ }\textbf {\bibinfo {volume}
  {33}},\ \bibinfo {pages} {6251} (\bibinfo {year} {2021})}\BibitemShut
  {NoStop}%
\bibitem [{\citenamefont {Irving}\ \emph {et~al.}(2017)\citenamefont {Irving},
  \citenamefont {Nicolini},\ and\ \citenamefont {Polcar}}]{Irving2017a}%
  \BibitemOpen
  \bibfield  {author} {\bibinfo {author} {\bibfnamefont {B.~J.}\ \bibnamefont
  {Irving}}, \bibinfo {author} {\bibfnamefont {P.}~\bibnamefont {Nicolini}},\
  and\ \bibinfo {author} {\bibfnamefont {T.}~\bibnamefont {Polcar}},\
  }\bibfield  {title} {\bibinfo {title} {{On the lubricity of transition metal
  dichalcogenides: an ab initio study}},\ }\href
  {https://doi.org/10.1039/C7NR00925A} {\bibfield  {journal} {\bibinfo
  {journal} {Nanoscale}\ }\textbf {\bibinfo {volume} {9}},\ \bibinfo {pages}
  {5597} (\bibinfo {year} {2017})}\BibitemShut {NoStop}%
\bibitem [{\citenamefont {Levita}\ \emph {et~al.}(2014)\citenamefont {Levita},
  \citenamefont {Cavaleiro}, \citenamefont {Molinari}, \citenamefont {Polcar},\
  and\ \citenamefont {Righi}}]{Levita2014a}%
  \BibitemOpen
  \bibfield  {author} {\bibinfo {author} {\bibfnamefont {G.}~\bibnamefont
  {Levita}}, \bibinfo {author} {\bibfnamefont {A.}~\bibnamefont {Cavaleiro}},
  \bibinfo {author} {\bibfnamefont {E.}~\bibnamefont {Molinari}}, \bibinfo
  {author} {\bibfnamefont {T.}~\bibnamefont {Polcar}},\ and\ \bibinfo {author}
  {\bibfnamefont {M.~C.}\ \bibnamefont {Righi}},\ }\bibfield  {title} {\bibinfo
  {title} {{Sliding properties of MoS$_{2}$ layers: Load and interlayer
  orientation effects}},\ }\href {https://doi.org/10.1021/jp4098099} {\bibfield
   {journal} {\bibinfo  {journal} {Journal of Physical Chemistry C}\ }\textbf
  {\bibinfo {volume} {118}},\ \bibinfo {pages} {13809} (\bibinfo {year}
  {2014})}\BibitemShut {NoStop}%
\bibitem [{\citenamefont {Silva}\ \emph {et~al.}(2021)\citenamefont {Silva},
  \citenamefont {Polcar},\ and\ \citenamefont {Kramer}}]{Silva2020a}%
  \BibitemOpen
  \bibfield  {author} {\bibinfo {author} {\bibfnamefont {A.}~\bibnamefont
  {Silva}}, \bibinfo {author} {\bibfnamefont {T.}~\bibnamefont {Polcar}},\ and\
  \bibinfo {author} {\bibfnamefont {D.}~\bibnamefont {Kramer}},\ }\bibfield
  {title} {\bibinfo {title} {{Phase behaviour of (Ti:Mo)S$_{2}$ binary alloys
  arising from electron-lattice coupling}},\ }\href
  {https://doi.org/10.1016/j.commatsci.2020.110044} {\bibfield  {journal}
  {\bibinfo  {journal} {Computational Materials Science}\ }\textbf {\bibinfo
  {volume} {186}},\ \bibinfo {pages} {110044} (\bibinfo {year}
  {2021})}\BibitemShut {NoStop}%
\bibitem [{\citenamefont {Jain}\ \emph {et~al.}(2013)\citenamefont {Jain},
  \citenamefont {Ong}, \citenamefont {Hautier}, \citenamefont {Chen},
  \citenamefont {Richards}, \citenamefont {Dacek}, \citenamefont {Cholia},
  \citenamefont {Gunter}, \citenamefont {Skinner}, \citenamefont {Ceder},\ and\
  \citenamefont {Persson}}]{Jain2013a}%
  \BibitemOpen
  \bibfield  {author} {\bibinfo {author} {\bibfnamefont {A.}~\bibnamefont
  {Jain}}, \bibinfo {author} {\bibfnamefont {S.~P.}\ \bibnamefont {Ong}},
  \bibinfo {author} {\bibfnamefont {G.}~\bibnamefont {Hautier}}, \bibinfo
  {author} {\bibfnamefont {W.}~\bibnamefont {Chen}}, \bibinfo {author}
  {\bibfnamefont {W.~D.}\ \bibnamefont {Richards}}, \bibinfo {author}
  {\bibfnamefont {S.}~\bibnamefont {Dacek}}, \bibinfo {author} {\bibfnamefont
  {S.}~\bibnamefont {Cholia}}, \bibinfo {author} {\bibfnamefont
  {D.}~\bibnamefont {Gunter}}, \bibinfo {author} {\bibfnamefont
  {D.}~\bibnamefont {Skinner}}, \bibinfo {author} {\bibfnamefont
  {G.}~\bibnamefont {Ceder}},\ and\ \bibinfo {author} {\bibfnamefont {K.~A.}\
  \bibnamefont {Persson}},\ }\bibfield  {title} {\bibinfo {title} {{Commentary:
  The materials project: A materials genome approach to accelerating materials
  innovation}},\ }\href {https://doi.org/10.1063/1.4812323} {\bibfield
  {journal} {\bibinfo  {journal} {APL Materials}\ }\textbf {\bibinfo {volume}
  {1}} (\bibinfo {year} {2013})}\BibitemShut {NoStop}%
\bibitem [{\citenamefont {Ong}\ \emph {et~al.}(2008)\citenamefont {Ong},
  \citenamefont {Wang}, \citenamefont {Kang},\ and\ \citenamefont
  {Ceder}}]{Ong2008}%
  \BibitemOpen
  \bibfield  {author} {\bibinfo {author} {\bibfnamefont {S.~P.}\ \bibnamefont
  {Ong}}, \bibinfo {author} {\bibfnamefont {L.}~\bibnamefont {Wang}}, \bibinfo
  {author} {\bibfnamefont {B.}~\bibnamefont {Kang}},\ and\ \bibinfo {author}
  {\bibfnamefont {G.}~\bibnamefont {Ceder}},\ }\bibfield  {title} {\bibinfo
  {title} {{Li - Fe - P - O$_{2}$ phase diagram from first principles
  calculations}},\ }\href {https://doi.org/10.1021/cm702327g} {\bibfield
  {journal} {\bibinfo  {journal} {Chemistry of Materials}\ }\textbf {\bibinfo
  {volume} {20}},\ \bibinfo {pages} {1798} (\bibinfo {year}
  {2008})}\BibitemShut {NoStop}%
\bibitem [{\citenamefont {Mar}\ \emph {et~al.}(1992)\citenamefont {Mar},
  \citenamefont {Jobic},\ and\ \citenamefont {Ibers}}]{Mar92WTe2}%
  \BibitemOpen
  \bibfield  {author} {\bibinfo {author} {\bibfnamefont {A.}~\bibnamefont
  {Mar}}, \bibinfo {author} {\bibfnamefont {S.}~\bibnamefont {Jobic}},\ and\
  \bibinfo {author} {\bibfnamefont {J.~A.}\ \bibnamefont {Ibers}},\ }\bibfield
  {title} {\bibinfo {title} {Metal-metal vs tellurium-tellurium bonding in
  {WTe}$_{2}$ and its ternary variants {TaIrTe}$_{4}$ and {NbIrTe}$_{4}$},\
  }\href {https://doi.org/10.1021/ja00049a029} {\bibfield  {journal} {\bibinfo
  {journal} {Journal of the American Chemical Society}\ }\textbf {\bibinfo
  {volume} {114}},\ \bibinfo {pages} {8963} (\bibinfo {year}
  {1992})}\BibitemShut {NoStop}%
\bibitem [{\citenamefont {Yu}\ \emph {et~al.}(2015)\citenamefont {Yu},
  \citenamefont {Aykol},\ and\ \citenamefont {Wolverton}}]{Yu2015thermoChem}%
  \BibitemOpen
  \bibfield  {author} {\bibinfo {author} {\bibfnamefont {Y.}~\bibnamefont
  {Yu}}, \bibinfo {author} {\bibfnamefont {M.}~\bibnamefont {Aykol}},\ and\
  \bibinfo {author} {\bibfnamefont {C.}~\bibnamefont {Wolverton}},\ }\bibfield
  {title} {\bibinfo {title} {{Reaction thermochemistry of metal sulfides with
  GGA and GGA+U calculations}},\ }\href
  {https://doi.org/10.1103/PhysRevB.92.195118} {\bibfield  {journal} {\bibinfo
  {journal} {Physical Review B}\ }\textbf {\bibinfo {volume} {92}} (\bibinfo
  {year} {2015})}\BibitemShut {NoStop}%
\bibitem [{\citenamefont {Ford}(2013)}]{ford2013statistical}%
  \BibitemOpen
  \bibfield  {author} {\bibinfo {author} {\bibfnamefont {I.}~\bibnamefont
  {Ford}},\ }\href@noop {} {\emph {\bibinfo {title} {{Statistical Physics: an
  entropic approach}}}}\ (\bibinfo  {publisher} {John Wiley {\&} Sons},\
  \bibinfo {year} {2013})\BibitemShut {NoStop}%
\bibitem [{\citenamefont {Sun}\ \emph {et~al.}(2016)\citenamefont {Sun},
  \citenamefont {Dacek}, \citenamefont {Ong}, \citenamefont {Hautier},
  \citenamefont {Jain}, \citenamefont {Richards}, \citenamefont {Gamst},
  \citenamefont {Persson},\ and\ \citenamefont {Ceder}}]{Sun2016a}%
  \BibitemOpen
  \bibfield  {author} {\bibinfo {author} {\bibfnamefont {W.}~\bibnamefont
  {Sun}}, \bibinfo {author} {\bibfnamefont {S.~T.}\ \bibnamefont {Dacek}},
  \bibinfo {author} {\bibfnamefont {S.~P.}\ \bibnamefont {Ong}}, \bibinfo
  {author} {\bibfnamefont {G.}~\bibnamefont {Hautier}}, \bibinfo {author}
  {\bibfnamefont {A.}~\bibnamefont {Jain}}, \bibinfo {author} {\bibfnamefont
  {W.~D.}\ \bibnamefont {Richards}}, \bibinfo {author} {\bibfnamefont {A.~C.}\
  \bibnamefont {Gamst}}, \bibinfo {author} {\bibfnamefont {K.~A.}\ \bibnamefont
  {Persson}},\ and\ \bibinfo {author} {\bibfnamefont {G.}~\bibnamefont
  {Ceder}},\ }\bibfield  {title} {\bibinfo {title} {{The thermodynamic scale of
  inorganic crystalline metastability}},\ }\href
  {https://www.science.org/doi/10.1126/sciadv.1600225} {\bibfield  {journal}
  {\bibinfo  {journal} {Science Advances}\ }\textbf {\bibinfo {volume} {2}}
  (\bibinfo {year} {2016})}\BibitemShut {NoStop}%
\bibitem [{\citenamefont {Zhu}\ \emph {et~al.}(2019)\citenamefont {Zhu},
  \citenamefont {Li}, \citenamefont {Inomata}, \citenamefont {Toda},\ and\
  \citenamefont {Ono}}]{Zhu2019d}%
  \BibitemOpen
  \bibfield  {author} {\bibinfo {author} {\bibfnamefont {M.}~\bibnamefont
  {Zhu}}, \bibinfo {author} {\bibfnamefont {J.}~\bibnamefont {Li}}, \bibinfo
  {author} {\bibfnamefont {N.}~\bibnamefont {Inomata}}, \bibinfo {author}
  {\bibfnamefont {M.}~\bibnamefont {Toda}},\ and\ \bibinfo {author}
  {\bibfnamefont {T.}~\bibnamefont {Ono}},\ }\bibfield  {title} {\bibinfo
  {title} {{Vanadium-doped molybdenum disulfide film-based strain sensors with
  high gauge factor}},\ }\href {https://doi.org/10.7567/1882-0786/aaf5c4}
  {\bibfield  {journal} {\bibinfo  {journal} {Applied Physics Express}\
  }\textbf {\bibinfo {volume} {12}},\ \bibinfo {pages} {015003} (\bibinfo
  {year} {2019})}\BibitemShut {NoStop}%
\bibitem [{\citenamefont {Xia}\ \emph {et~al.}(2021)\citenamefont {Xia},
  \citenamefont {Loh}, \citenamefont {Viner}, \citenamefont {Teutsch},
  \citenamefont {Graham}, \citenamefont {Kandyba}, \citenamefont {Barinov},
  \citenamefont {Sanchez}, \citenamefont {Smith}, \citenamefont {Hine},\ and\
  \citenamefont {Wilson}}]{Xia2021}%
  \BibitemOpen
  \bibfield  {author} {\bibinfo {author} {\bibfnamefont {X.}~\bibnamefont
  {Xia}}, \bibinfo {author} {\bibfnamefont {S.~M.}\ \bibnamefont {Loh}},
  \bibinfo {author} {\bibfnamefont {J.}~\bibnamefont {Viner}}, \bibinfo
  {author} {\bibfnamefont {N.~C.}\ \bibnamefont {Teutsch}}, \bibinfo {author}
  {\bibfnamefont {A.~J.}\ \bibnamefont {Graham}}, \bibinfo {author}
  {\bibfnamefont {V.}~\bibnamefont {Kandyba}}, \bibinfo {author} {\bibfnamefont
  {A.}~\bibnamefont {Barinov}}, \bibinfo {author} {\bibfnamefont {A.~M.}\
  \bibnamefont {Sanchez}}, \bibinfo {author} {\bibfnamefont {D.~C.}\
  \bibnamefont {Smith}}, \bibinfo {author} {\bibfnamefont {N.~D.}\ \bibnamefont
  {Hine}},\ and\ \bibinfo {author} {\bibfnamefont {N.~R.}\ \bibnamefont
  {Wilson}},\ }\bibfield  {title} {\bibinfo {title} {Atomic and electronic
  structure of two-dimensional {Mo}$_{(1-x)}${W}$_{x}${S}$_{2}$ alloys},\
  }\href {https://doi.org/10.1088/2515-7639/abdc6e} {\bibfield  {journal}
  {\bibinfo  {journal} {J. Phys. Mater.}\ }\textbf {\bibinfo {volume} {4}}
  (\bibinfo {year} {2021})}\BibitemShut {NoStop}%
\bibitem [{\citenamefont {Stolz}\ \emph {et~al.}(2022)\citenamefont {Stolz},
  \citenamefont {Kozhakhmetov}, \citenamefont {Dong}, \citenamefont
  {Gr{\"{o}}ning}, \citenamefont {Robinson},\ and\ \citenamefont
  {Schuler}}]{Stolz2022VWSe2_alloy}%
  \BibitemOpen
  \bibfield  {author} {\bibinfo {author} {\bibfnamefont {S.}~\bibnamefont
  {Stolz}}, \bibinfo {author} {\bibfnamefont {A.}~\bibnamefont {Kozhakhmetov}},
  \bibinfo {author} {\bibfnamefont {C.}~\bibnamefont {Dong}}, \bibinfo {author}
  {\bibfnamefont {O.}~\bibnamefont {Gr{\"{o}}ning}}, \bibinfo {author}
  {\bibfnamefont {J.~A.}\ \bibnamefont {Robinson}},\ and\ \bibinfo {author}
  {\bibfnamefont {B.}~\bibnamefont {Schuler}},\ }\bibfield  {title} {\bibinfo
  {title} {Layer-dependent schottky contact at van der waals interfaces:
  V-doped {WSe}$_{2}$ on graphene},\ }\href
  {https://doi.org/10.1038/s41699-022-00342-4} {\bibfield  {journal} {\bibinfo
  {journal} {npj 2D Materials and Applications}\ }\textbf {\bibinfo {volume}
  {6}},\ \bibinfo {pages} {66} (\bibinfo {year} {2022})}\BibitemShut {NoStop}%
\bibitem [{\citenamefont {Ahmad}\ \emph {et~al.}(2019)\citenamefont {Ahmad},
  \citenamefont {Reduan}, \citenamefont {Aidit}, \citenamefont {Yusoff},
  \citenamefont {Maah}, \citenamefont {Ismail},\ and\ \citenamefont
  {Tiu}}]{Ahmad2019MoWSe_alloy}%
  \BibitemOpen
  \bibfield  {author} {\bibinfo {author} {\bibfnamefont {H.}~\bibnamefont
  {Ahmad}}, \bibinfo {author} {\bibfnamefont {S.~A.}\ \bibnamefont {Reduan}},
  \bibinfo {author} {\bibfnamefont {S.~N.}\ \bibnamefont {Aidit}}, \bibinfo
  {author} {\bibfnamefont {N.}~\bibnamefont {Yusoff}}, \bibinfo {author}
  {\bibfnamefont {M.~J.}\ \bibnamefont {Maah}}, \bibinfo {author}
  {\bibfnamefont {M.~F.}\ \bibnamefont {Ismail}},\ and\ \bibinfo {author}
  {\bibfnamefont {Z.~C.}\ \bibnamefont {Tiu}},\ }\bibfield  {title} {\bibinfo
  {title} {Ternary {MoWSe}$_{2}$ alloy saturable absorber for passively
  q-switched yb-, er- and tm-doped fiber laser},\ }\href
  {https://doi.org/10.1016/j.optcom.2019.01.009} {\bibfield  {journal}
  {\bibinfo  {journal} {Optics Communications}\ }\textbf {\bibinfo {volume}
  {437}},\ \bibinfo {pages} {355} (\bibinfo {year} {2019})}\BibitemShut
  {NoStop}%
\bibitem [{\citenamefont {Susarla}\ \emph {et~al.}(2017)\citenamefont
  {Susarla}, \citenamefont {Kutana}, \citenamefont {Hachtel}, \citenamefont
  {Kochat}, \citenamefont {Apte}, \citenamefont {Vajtai}, \citenamefont
  {Idrobo}, \citenamefont {Yakobson}, \citenamefont {Tiwary},\ and\
  \citenamefont {Ajayan}}]{Susarla2017MoWSSe_alloy}%
  \BibitemOpen
  \bibfield  {author} {\bibinfo {author} {\bibfnamefont {S.}~\bibnamefont
  {Susarla}}, \bibinfo {author} {\bibfnamefont {A.}~\bibnamefont {Kutana}},
  \bibinfo {author} {\bibfnamefont {J.~A.}\ \bibnamefont {Hachtel}}, \bibinfo
  {author} {\bibfnamefont {V.}~\bibnamefont {Kochat}}, \bibinfo {author}
  {\bibfnamefont {A.}~\bibnamefont {Apte}}, \bibinfo {author} {\bibfnamefont
  {R.}~\bibnamefont {Vajtai}}, \bibinfo {author} {\bibfnamefont {J.~C.}\
  \bibnamefont {Idrobo}}, \bibinfo {author} {\bibfnamefont {B.~I.}\
  \bibnamefont {Yakobson}}, \bibinfo {author} {\bibfnamefont {C.~S.}\
  \bibnamefont {Tiwary}},\ and\ \bibinfo {author} {\bibfnamefont {P.~M.}\
  \bibnamefont {Ajayan}},\ }\bibfield  {title} {\bibinfo {title} {{Quaternary
  2D Transition Metal Dichalcogenides (TMDs) with Tunable Bandgap}},\ }\href
  {https://doi.org/10.1002/adma.201702457} {\bibfield  {journal} {\bibinfo
  {journal} {Advanced Materials}\ }\textbf {\bibinfo {volume} {29}} (\bibinfo
  {year} {2017})}\BibitemShut {NoStop}%
\bibitem [{\citenamefont {Su}\ \emph {et~al.}(2014)\citenamefont {Su},
  \citenamefont {Hsu}, \citenamefont {Chang}, \citenamefont {Chiu},
  \citenamefont {Hsu}, \citenamefont {Hsu}, \citenamefont {Chang},
  \citenamefont {He},\ and\ \citenamefont {Li}}]{Su2014MoSSe_alloy}%
  \BibitemOpen
  \bibfield  {author} {\bibinfo {author} {\bibfnamefont {S.~H.}\ \bibnamefont
  {Su}}, \bibinfo {author} {\bibfnamefont {Y.~T.}\ \bibnamefont {Hsu}},
  \bibinfo {author} {\bibfnamefont {Y.~H.}\ \bibnamefont {Chang}}, \bibinfo
  {author} {\bibfnamefont {M.~H.}\ \bibnamefont {Chiu}}, \bibinfo {author}
  {\bibfnamefont {C.~L.}\ \bibnamefont {Hsu}}, \bibinfo {author} {\bibfnamefont
  {W.~T.}\ \bibnamefont {Hsu}}, \bibinfo {author} {\bibfnamefont {W.~H.}\
  \bibnamefont {Chang}}, \bibinfo {author} {\bibfnamefont {J.~H.}\ \bibnamefont
  {He}},\ and\ \bibinfo {author} {\bibfnamefont {L.~J.}\ \bibnamefont {Li}},\
  }\bibfield  {title} {\bibinfo {title} {{Band gap-tunable molybdenum sulfide
  selenide monolayer alloy}},\ }\href {https://doi.org/10.1002/smll.201302893}
  {\bibfield  {journal} {\bibinfo  {journal} {Small}\ }\textbf {\bibinfo
  {volume} {10}},\ \bibinfo {pages} {2589} (\bibinfo {year}
  {2014})}\BibitemShut {NoStop}%
\bibitem [{\citenamefont {Liu}\ \emph {et~al.}(2021)\citenamefont {Liu},
  \citenamefont {Feng}, \citenamefont {Cai}, \citenamefont {Liu}, \citenamefont
  {Li}, \citenamefont {Amjadian}, \citenamefont {Cai}, \citenamefont {Wong},
  \citenamefont {Tamtaji}, \citenamefont {An}, \citenamefont {Zhang},
  \citenamefont {Chen}, \citenamefont {Wang}, \citenamefont {Xu},\ and\
  \citenamefont {Luo}}]{Liu2021MoTe}%
  \BibitemOpen
  \bibfield  {author} {\bibinfo {author} {\bibfnamefont {Z.}~\bibnamefont
  {Liu}}, \bibinfo {author} {\bibfnamefont {S.}~\bibnamefont {Feng}}, \bibinfo
  {author} {\bibfnamefont {X.}~\bibnamefont {Cai}}, \bibinfo {author}
  {\bibfnamefont {H.}~\bibnamefont {Liu}}, \bibinfo {author} {\bibfnamefont
  {J.}~\bibnamefont {Li}}, \bibinfo {author} {\bibfnamefont {M.}~\bibnamefont
  {Amjadian}}, \bibinfo {author} {\bibfnamefont {Y.}~\bibnamefont {Cai}},
  \bibinfo {author} {\bibfnamefont {H.}~\bibnamefont {Wong}}, \bibinfo {author}
  {\bibfnamefont {M.}~\bibnamefont {Tamtaji}}, \bibinfo {author} {\bibfnamefont
  {L.}~\bibnamefont {An}}, \bibinfo {author} {\bibfnamefont {K.}~\bibnamefont
  {Zhang}}, \bibinfo {author} {\bibfnamefont {G.}~\bibnamefont {Chen}},
  \bibinfo {author} {\bibfnamefont {N.}~\bibnamefont {Wang}}, \bibinfo {author}
  {\bibfnamefont {Z.}~\bibnamefont {Xu}},\ and\ \bibinfo {author}
  {\bibfnamefont {Z.}~\bibnamefont {Luo}},\ }\bibfield  {title} {\bibinfo
  {title} {Large-size superlattices synthesized by sequential sulfur
  substitution-induced transformation of metastable {MoTe}$_{2}$},\ }\href
  {https://doi.org/10.1021/acs.chemmater.1c03663} {\bibfield  {journal}
  {\bibinfo  {journal} {Chemistry of Materials}\ }\textbf {\bibinfo {volume}
  {33}},\ \bibinfo {pages} {9760} (\bibinfo {year} {2021})}\BibitemShut
  {NoStop}%
\bibitem [{\citenamefont {Tang}\ \emph {et~al.}(2021)\citenamefont {Tang},
  \citenamefont {Shu}, \citenamefont {Yang}, \citenamefont {Zhang},
  \citenamefont {Sheng}, \citenamefont {Liang}, \citenamefont {Cao},\ and\
  \citenamefont {Chen}}]{Tang2021MoSSe_alloy}%
  \BibitemOpen
  \bibfield  {author} {\bibinfo {author} {\bibfnamefont {P.}~\bibnamefont
  {Tang}}, \bibinfo {author} {\bibfnamefont {H.}~\bibnamefont {Shu}}, \bibinfo
  {author} {\bibfnamefont {M.}~\bibnamefont {Yang}}, \bibinfo {author}
  {\bibfnamefont {M.}~\bibnamefont {Zhang}}, \bibinfo {author} {\bibfnamefont
  {C.}~\bibnamefont {Sheng}}, \bibinfo {author} {\bibfnamefont
  {P.}~\bibnamefont {Liang}}, \bibinfo {author} {\bibfnamefont
  {D.}~\bibnamefont {Cao}},\ and\ \bibinfo {author} {\bibfnamefont
  {X.}~\bibnamefont {Chen}},\ }\bibfield  {title} {\bibinfo {title} {{Rapid
  Wafer-Scale Growth of {MoS}$_{2(1- x)}${Se}$_{2x}$ Alloy Monolayers with
  Tunable Compositions and Optical Properties for High-Performance
  Photodetectors}},\ }\href {https://doi.org/10.1021/acsanm.1c03137} {\bibfield
   {journal} {\bibinfo  {journal} {ACS Applied Nano Materials}\ }\textbf
  {\bibinfo {volume} {4}},\ \bibinfo {pages} {12609} (\bibinfo {year}
  {2021})}\BibitemShut {NoStop}%
\bibitem [{\citenamefont {Kim}\ \emph {et~al.}(2021)\citenamefont {Kim},
  \citenamefont {Seung}, \citenamefont {Kang}, \citenamefont {Kim},
  \citenamefont {Bae}, \citenamefont {Park}, \citenamefont {Kang},
  \citenamefont {Choi}, \citenamefont {Choi}, \citenamefont {Kim},
  \citenamefont {Hyeon}, \citenamefont {Lee}, \citenamefont {Kim},
  \citenamefont {Shim},\ and\ \citenamefont {Park}}]{Kim2021TMD_alloy}%
  \BibitemOpen
  \bibfield  {author} {\bibinfo {author} {\bibfnamefont {J.}~\bibnamefont
  {Kim}}, \bibinfo {author} {\bibfnamefont {H.}~\bibnamefont {Seung}}, \bibinfo
  {author} {\bibfnamefont {D.}~\bibnamefont {Kang}}, \bibinfo {author}
  {\bibfnamefont {J.}~\bibnamefont {Kim}}, \bibinfo {author} {\bibfnamefont
  {H.}~\bibnamefont {Bae}}, \bibinfo {author} {\bibfnamefont {H.}~\bibnamefont
  {Park}}, \bibinfo {author} {\bibfnamefont {S.}~\bibnamefont {Kang}}, \bibinfo
  {author} {\bibfnamefont {C.}~\bibnamefont {Choi}}, \bibinfo {author}
  {\bibfnamefont {B.~K.}\ \bibnamefont {Choi}}, \bibinfo {author}
  {\bibfnamefont {J.~S.}\ \bibnamefont {Kim}}, \bibinfo {author} {\bibfnamefont
  {T.}~\bibnamefont {Hyeon}}, \bibinfo {author} {\bibfnamefont
  {H.}~\bibnamefont {Lee}}, \bibinfo {author} {\bibfnamefont {D.~H.}\
  \bibnamefont {Kim}}, \bibinfo {author} {\bibfnamefont {S.}~\bibnamefont
  {Shim}},\ and\ \bibinfo {author} {\bibfnamefont {J.}~\bibnamefont {Park}},\
  }\bibfield  {title} {\bibinfo {title} {{Wafer-Scale Production of Transition
  Metal Dichalcogenides and Alloy Monolayers by Nanocrystal Conversion for
  Large-Scale Ultrathin Flexible Electronics}},\ }\href
  {https://doi.org/10.1021/acs.nanolett.1c02991} {\bibfield  {journal}
  {\bibinfo  {journal} {Nano Letters}\ }\textbf {\bibinfo {volume} {21}},\
  \bibinfo {pages} {9153} (\bibinfo {year} {2021})}\BibitemShut {NoStop}%
\bibitem [{\citenamefont {Yu}\ \emph {et~al.}(2017)\citenamefont {Yu},
  \citenamefont {Lin}, \citenamefont {Sun}, \citenamefont {Le}, \citenamefont
  {Yu}, \citenamefont {Gao}, \citenamefont {Hsu}, \citenamefont {Wu},
  \citenamefont {Chang}, \citenamefont {Zeng}, \citenamefont {Liu},
  \citenamefont {Wang}, \citenamefont {Jeng}, \citenamefont {Lin},
  \citenamefont {Trampert}, \citenamefont {Shen}, \citenamefont {Suenaga},\
  and\ \citenamefont {Liu}}]{Yu2016WSe-Te2_phaseT}%
  \BibitemOpen
  \bibfield  {author} {\bibinfo {author} {\bibfnamefont {P.}~\bibnamefont
  {Yu}}, \bibinfo {author} {\bibfnamefont {J.}~\bibnamefont {Lin}}, \bibinfo
  {author} {\bibfnamefont {L.}~\bibnamefont {Sun}}, \bibinfo {author}
  {\bibfnamefont {Q.~L.}\ \bibnamefont {Le}}, \bibinfo {author} {\bibfnamefont
  {X.}~\bibnamefont {Yu}}, \bibinfo {author} {\bibfnamefont {G.}~\bibnamefont
  {Gao}}, \bibinfo {author} {\bibfnamefont {C.~H.}\ \bibnamefont {Hsu}},
  \bibinfo {author} {\bibfnamefont {D.}~\bibnamefont {Wu}}, \bibinfo {author}
  {\bibfnamefont {T.~R.}\ \bibnamefont {Chang}}, \bibinfo {author}
  {\bibfnamefont {Q.}~\bibnamefont {Zeng}}, \bibinfo {author} {\bibfnamefont
  {F.}~\bibnamefont {Liu}}, \bibinfo {author} {\bibfnamefont {Q.~J.}\
  \bibnamefont {Wang}}, \bibinfo {author} {\bibfnamefont {H.~T.}\ \bibnamefont
  {Jeng}}, \bibinfo {author} {\bibfnamefont {H.}~\bibnamefont {Lin}}, \bibinfo
  {author} {\bibfnamefont {A.}~\bibnamefont {Trampert}}, \bibinfo {author}
  {\bibfnamefont {Z.}~\bibnamefont {Shen}}, \bibinfo {author} {\bibfnamefont
  {K.}~\bibnamefont {Suenaga}},\ and\ \bibinfo {author} {\bibfnamefont
  {Z.}~\bibnamefont {Liu}},\ }\bibfield  {title} {\bibinfo {title}
  {Metal-semiconductor phase-transition in {WSe}$_{2(1-x)}${Te}$_{2x}$
  monolayer},\ }\href {https://doi.org/10.1002/adma.201603991} {\bibfield
  {journal} {\bibinfo  {journal} {Advanced Materials}\ }\textbf {\bibinfo
  {volume} {29}} (\bibinfo {year} {2017})}\BibitemShut {NoStop}%
\bibitem [{\citenamefont {Van~der Ven}\ \emph {et~al.}(2010)\citenamefont
  {Van~der Ven}, \citenamefont {Thomas}, \citenamefont {Xu},\ and\
  \citenamefont {Bhattacharya}}]{VanderVen2010}%
  \BibitemOpen
  \bibfield  {author} {\bibinfo {author} {\bibfnamefont {A.}~\bibnamefont
  {Van~der Ven}}, \bibinfo {author} {\bibfnamefont {J.~C.}\ \bibnamefont
  {Thomas}}, \bibinfo {author} {\bibfnamefont {Q.}~\bibnamefont {Xu}},\ and\
  \bibinfo {author} {\bibfnamefont {J.}~\bibnamefont {Bhattacharya}},\
  }\bibfield  {title} {\bibinfo {title} {Linking the electronic structure of
  solids to their thermodynamic and kinetic properties},\ }\href
  {https://doi.org/10.1016/j.matcom.2009.08.008} {\bibfield  {journal}
  {\bibinfo  {journal} {Mathematics and Computers in Simulation}\ }\textbf
  {\bibinfo {volume} {80}},\ \bibinfo {pages} {1393} (\bibinfo {year}
  {2010})}\BibitemShut {NoStop}%
\bibitem [{\citenamefont {Puchala}\ and\ \citenamefont {Van
  Der~Ven}(2013)}]{Puchala2013}%
  \BibitemOpen
  \bibfield  {author} {\bibinfo {author} {\bibfnamefont {B.}~\bibnamefont
  {Puchala}}\ and\ \bibinfo {author} {\bibfnamefont {A.}~\bibnamefont {Van
  Der~Ven}},\ }\bibfield  {title} {\bibinfo {title} {{Thermodynamics of the
  Zr-O system from first-principles calculations}},\ }\href
  {https://doi.org/10.1103/PhysRevB.88.094108} {\bibfield  {journal} {\bibinfo
  {journal} {Physical Review B}\ }\textbf {\bibinfo {volume} {88}},\ \bibinfo
  {pages} {1} (\bibinfo {year} {2013})}\BibitemShut {NoStop}%
\bibitem [{\citenamefont {Thomas}\ and\ \citenamefont
  {Ven}(2013)}]{Thomas2013}%
  \BibitemOpen
  \bibfield  {author} {\bibinfo {author} {\bibfnamefont {J.~C.}\ \bibnamefont
  {Thomas}}\ and\ \bibinfo {author} {\bibfnamefont {A.~V.~D.}\ \bibnamefont
  {Ven}},\ }\bibfield  {title} {\bibinfo {title} {{Finite-temperature
  properties of strongly anharmonic and mechanically unstable crystal phases
  from first principles}},\ }\href {https://doi.org/10.1103/PhysRevB.88.214111}
  {\bibfield  {journal} {\bibinfo  {journal} {Physical Review B}\ }\textbf
  {\bibinfo {volume} {88}},\ \bibinfo {pages} {1} (\bibinfo {year}
  {2013})}\BibitemShut {NoStop}%
\bibitem [{\citenamefont {Thomas}\ \emph {et~al.}(2021)\citenamefont {Thomas},
  \citenamefont {Natarajan},\ and\ \citenamefont {Van~der
  Ven}}]{Thomas2021ComparingGeometry}%
  \BibitemOpen
  \bibfield  {author} {\bibinfo {author} {\bibfnamefont {J.~C.}\ \bibnamefont
  {Thomas}}, \bibinfo {author} {\bibfnamefont {A.~R.}\ \bibnamefont
  {Natarajan}},\ and\ \bibinfo {author} {\bibfnamefont {A.}~\bibnamefont
  {Van~der Ven}},\ }\bibfield  {title} {\bibinfo {title} {{Comparing crystal
  structures with symmetry and geometry}},\ }\href
  {https://doi.org/10.1038/s41524-021-00627-0} {\bibfield  {journal} {\bibinfo
  {journal} {npj Computational Materials}\ }\textbf {\bibinfo {volume} {7}}
  (\bibinfo {year} {2021})}\BibitemShut {NoStop}%
\bibitem [{\citenamefont {van~de Walle}\ and\ \citenamefont
  {Asta}(2002)}]{VandeWalle2002b}%
  \BibitemOpen
  \bibfield  {author} {\bibinfo {author} {\bibfnamefont {A.}~\bibnamefont
  {van~de Walle}}\ and\ \bibinfo {author} {\bibfnamefont {M.}~\bibnamefont
  {Asta}},\ }\bibfield  {title} {\bibinfo {title} {{Self-driven Lattice-model
  Monte Carlo Simulations of Alloy Thermodynamic Properties and Phase
  Diagrams}},\ }\href
  {http://iopscience.iop.org/article/10.1088/0965-0393/10/5/304/pdf} {\bibfield
   {journal} {\bibinfo  {journal} {Model. Simul. Mater. Sc.}\ }\textbf
  {\bibinfo {volume} {10}},\ \bibinfo {pages} {521} (\bibinfo {year}
  {2002})}\BibitemShut {NoStop}%
\bibitem [{\citenamefont {Hsu}\ \emph {et~al.}(2001)\citenamefont {Hsu},
  \citenamefont {Zhu}, \citenamefont {Yao}, \citenamefont {Firth},
  \citenamefont {Clark}, \citenamefont {Kroto},\ and\ \citenamefont
  {Walton}}]{Hsu2001}%
  \BibitemOpen
  \bibfield  {author} {\bibinfo {author} {\bibfnamefont {W.~K.}\ \bibnamefont
  {Hsu}}, \bibinfo {author} {\bibfnamefont {Y.~Q.}\ \bibnamefont {Zhu}},
  \bibinfo {author} {\bibfnamefont {N.}~\bibnamefont {Yao}}, \bibinfo {author}
  {\bibfnamefont {S.}~\bibnamefont {Firth}}, \bibinfo {author} {\bibfnamefont
  {R.~J.~H.}\ \bibnamefont {Clark}}, \bibinfo {author} {\bibfnamefont {H.~W.}\
  \bibnamefont {Kroto}},\ and\ \bibinfo {author} {\bibfnamefont {D.~R.~M.}\
  \bibnamefont {Walton}},\ }\bibfield  {title} {\bibinfo {title}
  {{Titanium-doped molybdenum disulfide nanostructures}},\ }\href
  {https://doi.org/10.1002/1616-3028(200102)11:1<69::AID-ADFM69>3.0.CO;2-D}
  {\bibfield  {journal} {\bibinfo  {journal} {Advanced Funtional Materials}\
  }\textbf {\bibinfo {volume} {11}},\ \bibinfo {pages} {69} (\bibinfo {year}
  {2001})}\BibitemShut {NoStop}%
\bibitem [{\citenamefont {Bart{\'{o}}k}\ \emph {et~al.}(2013)\citenamefont
  {Bart{\'{o}}k}, \citenamefont {Kondor},\ and\ \citenamefont
  {Cs{\'{a}}nyi}}]{Bartok2013SOAP}%
  \BibitemOpen
  \bibfield  {author} {\bibinfo {author} {\bibfnamefont {A.~P.}\ \bibnamefont
  {Bart{\'{o}}k}}, \bibinfo {author} {\bibfnamefont {R.}~\bibnamefont
  {Kondor}},\ and\ \bibinfo {author} {\bibfnamefont {G.}~\bibnamefont
  {Cs{\'{a}}nyi}},\ }\bibfield  {title} {\bibinfo {title} {On representing
  chemical environments},\ }\href {https://doi.org/10.1103/PhysRevB.87.184115}
  {\bibfield  {journal} {\bibinfo  {journal} {Physical Review B}\ }\textbf
  {\bibinfo {volume} {87}} (\bibinfo {year} {2013})}\BibitemShut {NoStop}%
\bibitem [{\citenamefont {De}\ \emph {et~al.}(2016)\citenamefont {De},
  \citenamefont {Bart{\'{o}}k}, \citenamefont {Cs{\'{a}}nyi},\ and\
  \citenamefont {Ceriotti}}]{Bartok2016SOAP}%
  \BibitemOpen
  \bibfield  {author} {\bibinfo {author} {\bibfnamefont {S.}~\bibnamefont
  {De}}, \bibinfo {author} {\bibfnamefont {A.~P.}\ \bibnamefont
  {Bart{\'{o}}k}}, \bibinfo {author} {\bibfnamefont {G.}~\bibnamefont
  {Cs{\'{a}}nyi}},\ and\ \bibinfo {author} {\bibfnamefont {M.}~\bibnamefont
  {Ceriotti}},\ }\bibfield  {title} {\bibinfo {title} {{Comparing molecules and
  solids across structural and alchemical space}},\ }\href
  {https://doi.org/10.1039/c6cp00415f} {\bibfield  {journal} {\bibinfo
  {journal} {Physical Chemistry Chemical Physics}\ }\textbf {\bibinfo {volume}
  {18}},\ \bibinfo {pages} {13754} (\bibinfo {year} {2016})}\BibitemShut
  {NoStop}%
\bibitem [{\citenamefont {Thompson}\ \emph {et~al.}(1972)\citenamefont
  {Thompson}, \citenamefont {Pisharody},\ and\ \citenamefont
  {Koehler}}]{THOMPSON1972}%
  \BibitemOpen
  \bibfield  {author} {\bibinfo {author} {\bibfnamefont {A.~H.}\ \bibnamefont
  {Thompson}}, \bibinfo {author} {\bibfnamefont {K.~R.}\ \bibnamefont
  {Pisharody}},\ and\ \bibinfo {author} {\bibfnamefont {R.~F.}\ \bibnamefont
  {Koehler}},\ }\bibfield  {title} {\bibinfo {title} {Experimental study of the
  solid solutions {Ti}$_{x}${Ta}$_{1-x}${S}$_{2}$},\ }\href
  {https://doi.org/10.1103/PhysRevLett.29.163} {\bibfield  {journal} {\bibinfo
  {journal} {Physical Review Letters}\ }\textbf {\bibinfo {volume} {29}},\
  \bibinfo {pages} {163} (\bibinfo {year} {1972})}\BibitemShut {NoStop}%
\bibitem [{\citenamefont {Kresse}\ and\ \citenamefont
  {Furthm{\"{u}}ller}(1996)}]{KresseAv1996}%
  \BibitemOpen
  \bibfield  {author} {\bibinfo {author} {\bibfnamefont {G.}~\bibnamefont
  {Kresse}}\ and\ \bibinfo {author} {\bibfnamefont {J.}~\bibnamefont
  {Furthm{\"{u}}ller}},\ }\bibfield  {title} {\bibinfo {title} {{Efficiency of
  ab-initio total energy calculations for metals and semiconductors using a
  plane-wave basis set}},\ }\href
  {https://doi.org/10.1016/0927-0256(96)00008-0} {\bibfield  {journal}
  {\bibinfo  {journal} {Computational Materials Science}\ }\textbf {\bibinfo
  {volume} {6}},\ \bibinfo {pages} {15} (\bibinfo {year} {1996})}\BibitemShut
  {NoStop}%
\bibitem [{\citenamefont {Kresse}\ and\ \citenamefont
  {Hafner}(1993)}]{Kresse1993}%
  \BibitemOpen
  \bibfield  {author} {\bibinfo {author} {\bibfnamefont {G.}~\bibnamefont
  {Kresse}}\ and\ \bibinfo {author} {\bibfnamefont {J.}~\bibnamefont
  {Hafner}},\ }\bibfield  {title} {\bibinfo {title} {{Ab initio molecular
  dynamics for open-shell transition metals}},\ }\href
  {https://doi.org/10.1103/PhysRevB.48.13115} {\bibfield  {journal} {\bibinfo
  {journal} {Physical Review B}\ }\textbf {\bibinfo {volume} {48}},\ \bibinfo
  {pages} {13115} (\bibinfo {year} {1993})}\BibitemShut {NoStop}%
\bibitem [{\citenamefont {Kresse}\ and\ \citenamefont
  {Joubert}(1999)}]{Kresse1999}%
  \BibitemOpen
  \bibfield  {author} {\bibinfo {author} {\bibfnamefont {G.}~\bibnamefont
  {Kresse}}\ and\ \bibinfo {author} {\bibfnamefont {D.}~\bibnamefont
  {Joubert}},\ }\bibfield  {title} {\bibinfo {title} {{From ultrasoft
  pseudopotentials to the projector augmented-wave method}},\ }\href
  {https://doi.org/10.1103/PhysRevB.59.1758} {\bibfield  {journal} {\bibinfo
  {journal} {Physical Review B}\ }\textbf {\bibinfo {volume} {59}},\ \bibinfo
  {pages} {1758} (\bibinfo {year} {1999})}\BibitemShut {NoStop}%
\bibitem [{\citenamefont {Bl{\"{o}}chl}(1994)}]{Blochl1994}%
  \BibitemOpen
  \bibfield  {author} {\bibinfo {author} {\bibfnamefont {P.~E.}\ \bibnamefont
  {Bl{\"{o}}chl}},\ }\bibfield  {title} {\bibinfo {title} {{Projector
  augmented-wave method}},\ }\href {https://doi.org/10.1103/PhysRevB.50.17953}
  {\bibfield  {journal} {\bibinfo  {journal} {Physical Review B}\ }\textbf
  {\bibinfo {volume} {50}},\ \bibinfo {pages} {17953} (\bibinfo {year}
  {1994})}\BibitemShut {NoStop}%
\bibitem [{\citenamefont {Perdew}\ \emph {et~al.}(1996)\citenamefont {Perdew},
  \citenamefont {Burke},\ and\ \citenamefont {Ernzerhof}}]{Perdew1996}%
  \BibitemOpen
  \bibfield  {author} {\bibinfo {author} {\bibfnamefont {J.~P.}\ \bibnamefont
  {Perdew}}, \bibinfo {author} {\bibfnamefont {K.}~\bibnamefont {Burke}},\ and\
  \bibinfo {author} {\bibfnamefont {M.}~\bibnamefont {Ernzerhof}},\ }\bibfield
  {title} {\bibinfo {title} {{Generalized Gradient Approximation Made
  Simple}},\ }\href {https://doi.org/10.1103/PhysRevLett.77.3865} {\bibfield
  {journal} {\bibinfo  {journal} {Physical Review Letters}\ }\textbf {\bibinfo
  {volume} {77}},\ \bibinfo {pages} {3865} (\bibinfo {year}
  {1996})}\BibitemShut {NoStop}%
\end{thebibliography}
%\nocite{*}

\end{document}